\documentclass[prl,reprint,amsmath,amssymb,showpacs,superscriptaddress,twocolumn]{revtex4-1}

\usepackage[dvipdfmx]{graphicx}
\usepackage{bm}
\usepackage{mathrsfs}
\usepackage{color}

\DeclareMathOperator{\trace}{Tr}

\DeclareMathOperator{\diag}{diag}

\begin{document}
\title{Majorana Multipole Response of Topological Superconductors
%: Theory and application to half-Heusler superconductors
%: Theory and Application to superconductors with $J=\frac{3}{2}$ Fermions
}
\author {Shingo Kobayashi}
\affiliation{Institute for Advanced Research, Nagoya University, Nagoya 464-8601, Japan}
\affiliation{Department of Applied Physics, Nagoya University, Nagoya 464-8603, Japan}
\author{Ai Yamakage}
\affiliation{Department of Physics, Nagoya University, Nagoya 464-8602, Japan}
\author{Yukio Tanaka}
\affiliation{Department of Applied Physics, Nagoya University, Nagoya 464-8603, Japan}
\author{Masatoshi Sato}
\affiliation{Yukawa Institute for Theoretical Physics, Kyoto University, Kyoto 606-8502, Japan}

\date{\today}

\begin{abstract}
In contrast to elementary Majorana particles, emergent Majorana fermions (MFs) in condensed-matter systems may have electromagnetic multipoles. 
We developed a general theory of magnetic multipoles for surface helical MFs on time-reversal-invariant superconductors. 
The results show that the multipole response is governed by crystal symmetry, and that a one-to-one correspondence exists between the symmetry of Cooper pairs and the representation of magnetic multipoles under crystal symmetry. 
The latter property provides a way to identify nonconventional pairing symmetry via the magnetic response of surface MFs.
We also find that most helical MFs exhibit a magnetic-dipole response, but those on superconductors with spin-3/2 electrons may display a magnetic-octupole response in leading order, which uniquely characterizes high-spin superconductors.
Detection of such an octupole response provides direct evidence of high-spin superconductivity, such as in half-Heusler superconductors. 
\end{abstract}
\maketitle

\paragraph*{Introduction.} 
The emergence of Majorana fermions (MFs) in electron systems has led to intense interest in searching for such exotic new excitations in condensed-matter physics.
Particularly, recent developments have shown that
emergent MFs appear as gapless Andreev bound states in topological superconductors (TSCs)~\cite{Hu94,Kashiwaya2000,Volovik03, Schnyder08, Sato09, Wilczek09, 
Hasan10,Qi11,Tanaka12,Alicea12,SatoFujimoto16,Mizushima16,Chiu16,Sato17}, which provide a potential candidate for fault-tolerant qubits for topological quantum computation~\cite{Nayak08}.
The increased interest in topological materials has led to a proposal of versatile three-dimensional (3D) time-reversal-invariant (TR-invariant) TSCs, such as superconducting doped topological insulators (TIs)~\cite{Hor10,Fu10,Sasaki11,Sasaki12,Hashimoto15,Fu14,Matano16,Yonezawa16} and Dirac semimetals~\cite{Aggarwal15,Wang15,Kobayashi15PRL,Hashimoto16,Oudah16,Kawakami18}, which commonly host helical MFs forming Kramers pairs on their surfaces. 

Emergent MFs share some properties with elementary Majorana particles~\cite{Majorana37,Avignone08}. For example, both obey Dirac equations with charge-conjugation symmetry.
Furthermore, a pair of MF zero modes are required to define the fermionic creation and annihilation operators, from which zero modes exhibit non-Abelian anyon statistics. 
However, compared with elementary Majorana particles, emergent MFs respond very differently to electric and magnetic fields. 
Contrarily, neither electric nor magnetic multipoles are possible for elementary MFs ~\cite{Kayser83,Radescu85,Boudjema89}: 
$\mathcal{CPT}$ invariance, where $\mathcal{C}$ is charge conjugation, $\mathcal{P}$ is space inversion, and $\mathcal{T}$ is time reversal, is a fundamental symmetry that any relativistic elementary particles is expected to respect. 
This symmetry forbids intrinsic electric and magnetic multipoles for elementary Majorana particles because they are their own antiparticles under $\mathcal{CPT}$.
Contrarily, in superconductors, fundamental symmetry is just charge-conjugation (namely, particle-hole (PH) symmetry), and the emergent MFs are self-conjugate under $\mathcal{C}$.
Therefore, MFs in condensed-matter physics are not subject to such a strong constraint, and 
no systematic study on their electromagnetic multipoles has yet been attempted.

In this Letter, we develop a theory describing the electric and magnetic response of MFs in superconductors. 
For clarity, we focus here on surface helical MFs on 3D TR-invariant TSCs. 
A key ingredient specific to emergent MFs is crystalline symmetry. 
In analogy with $\mathcal{CPT}$ invariance for elementary MFs, crystal symmetry provides additional symmetry constraints on electromagnetic structures of emergent MFs. Considering the constraints, we establish a response theory for helical MFs in a low-energy limit, in which the problem reduces to the selection rule for crystal-symmetry groups. Applying our theory to possible crystal-symmetry groups, we find that helical MFs can host magnetic-multipole structures of dipole or octupole orders as the leading contribution. Additionally, the results predict a one-to-one correspondence between irreducible representation (IR) of Cooper pairs and magnetic multipoles, which helps to determine the pairing symmetry experimentally through the magnetic response of MFs.

Particularly, the proposed theory provides a unique way to identify topological superconductivity of spin-3/2 electrons.
Although research interest has recently focused on high-spin topological superconductivity~\cite{Goll08,Butch11,Tafti13,GXu16,Bay12,Kim18,Boettcher16,Brydon16,Savary17,Roy17,Timm17,Venderbos18,Boettcher18,Yu18,Kuzmenko18,Kawakami18}, 
little is known about distinguishing TSCs of spin-$3/2$ electrons from those of spin-$1/2$ electrons. 
Thus, we clarify that magnetic responses of helical MFs can unambiguously distinguish between these two types of SCs because the magnetic-octupole response is unique to higher-spin TSCs. 
To illustrate this, we apply the proposed theory to superconducting TIs of ordinary spin-1/2 electrons~\cite{Fu10} and parity-mixed half-Heusler superconductors of spin-$3/2$ electrons~\cite{Brydon16, Kim18}.
The results of both numerical and analytical analyses show that only the latter exhibits the octupole response under the same crystalline symmetry.

\paragraph*{Majorana multipole.} 
Helical MFs are a superconducting analogue of surface Dirac fermions of TIs and can be realized in 3D TR-invariant TSCs. From the bulk-boundary correspondence, the existence of helical MFs is ensured by the so-called 3D winding number~\cite{Grinevich88, Schnyder08, Sato09, Sato10}. Whereas the 3D winding number is defined only for fully gapped TSCs, its parity is well defined even for nodal superconductors~\cite{Sasaki11}. Provided TR symmetry is maintained, these invariants are well defined and protect surface helical MFs for both nodal and nodeless superconductors.

We consider the quantum response of helical MFs when exposed to external electric or magnetic fields. 
First, note that electric fields only elicit a moderate response from helical MFs because electric fields maintain TR symmetry, helical MFs remains gapless so their response is weak. Conversely, magnetic fields may substantially affect them. Magnetic fields break TR symmetry, so the 3D winding number and its parity become invalid.
However, this does not mean that helical MFs are not immune to some magnetic fields because actual TSCs have their own crystalline symmetry.
Depending on the direction of the applied magnetic field, 
TR symmetry may be partially preserved by combining it with crystalline symmetry. 
Such magnetic crystalline symmetry determines the stability of helical MFs under magnetic fields~\cite{note1}.

\begin{figure}[tbp]
\centering
 \includegraphics[width=6cm]{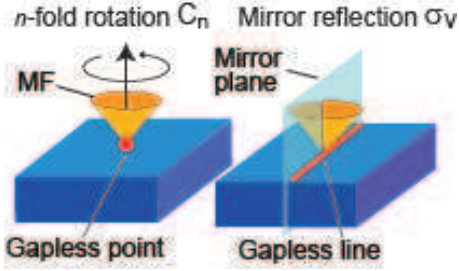}
 \caption{$n$-fold rotation and mirror reflection that are compatible with the surface.}\label{fig:mirror}
\end{figure}

As relevant point group operations, we now consider mirror reflections and rotations that are compatible with the surface in question.
The mirror plane and the rotation axis should be normal to the surface
(see Fig. \ref{fig:mirror}).
We consider all two-dimensional point groups formed by them: $C_2$, $C_3$, $C_4$, $C_6$, $C_s$, $C_{2v}$, $C_{3v}$, $C_{4v}$, and $C_{6v}$, in addition to TR symmetry.
Under a magnetic field,
we retain only magnetic mirror reflection (or magnetic two-fold rotation). 
Note that the retained magnetic symmetry is selected by the direction of an applied magnetic field: 
Only for a magnetic field parallel (normal) to the mirror plane (rotation axis) is magnetic mirror reflection (magnetic two-fold rotation) preserved. The above magnetic field is easily seen to flip under TR, but it points back to the original when we simultaneously do a mirror reflection (two-fold rotation).

The retained magnetic symmetry enables TSCs to host an additional topological number that is valid even when the TSC is exposed to a magnetic field.
Combining magnetic symmetry with PH symmetry, which is intrinsic to superconductors, one can introduce the magnetic one-dimensional (1D) winding number~\cite{Mizushima12, Shiozaki14,Dumitrescu2014,Xiong17}:
$
w_{\rm M1D}=\frac{i}{4\pi}\int dk_{\perp}{\rm tr}\left[\Gamma_{\rm M}{\cal H}^{-1}(k_{\perp},\bm{k}_{\parallel})
\partial_{k_{\perp}}{\cal H}(k_{\perp},\bm{k}_{\parallel})\right] 
$
where ${\cal H} (\bm{k})$ is the Bogoliubov-de Gennes (BdG) Hamiltonian, $(k_{\perp},\bm{k}_{\parallel})$ are the momentum normal and parallel to the surface, respectively, and $\Gamma_{\rm M} \equiv \mathcal{U} \mathcal{T} \mathcal{C}$ is the magnetic chiral operator. 
Here, $\mathcal{U}$ is a mirror reflection or two-fold rotation. 
If $w_{\rm M1D}$ for magnetic two-fold rotation (magnetic mirror reflection) is nonzero in the absence of magnetic fields, then helical MFs remain gapless even under a magnetic field normal (parallel) to the rotation axis (mirror plane), provided the system maintains the bulk gap.
Conversely, helical MFs do not necessarily remain gapless under other magnetic fields.
This direction dependence results in an anisotropic magnetic response of helical MFs.
Note that $w_{\rm M1D}$ for magnetic two-fold rotation (magnetic mirror reflection) is defined only on the symmetric axis (plane), so it protects the gapless point (line) of helical MFs at the symmetry axis (plane) in the surface Brillouin zone (see Fig. \ref{fig:mirror}). From the bulk-boundary correspondence, the gapless points or lines are obtained as zero modes $|u_0^{(a)}\rangle$ ($a=1,2,\dots$) of the BdG equation.

\begin{table*}[tb]
\caption{Magnetic multipole of MFs. From left to right, each column shows two-dimensional point groups (PGs), IRs of $\Delta_{\Gamma}$ with $w_{\rm M1D} \neq 0$, the basis of $\Delta_{\Gamma}$, $\mathcal{U}$ associated with $\Gamma_{\rm M} $, IRs of $\mathcal{O}_{\Gamma}$, and the basis of $\mathcal{O}_{\Gamma}$. Here, $J_i$ are the spin matrices, ``$-$'' means the absence of IRs, and $\mathcal{O}_{\Gamma}$ describes the leading order of the magnetic multipoles. 
}
\label{table}
\begin{tabular}{cccccc}
\hline\hline
PG & $\Delta_{\Gamma}$& basis of $\Delta_{\Gamma}$ $(\times e^{-i \pi J_y})$ &$\mathcal{U}$& ${\cal O}_{\Gamma}$ & basis of $\mathcal{O}_{\Gamma}$\\
\hline 
${\rm C_2}, {\rm C_4}, {\rm C_6}$ & A & $\bm{k} \cdot \bm{J}$& $C_2$  &A &$J_z$\\
${\rm C_3}$ & $-$ &$-$ &$-$&$-$& $-$\\
%${\rm C_4}$ & A & $\bm{k} \cdot \bm{J}$ & $C_2$&A &$J_z$\\
% &   & & &(${\rm B}$)&$(J_xJ_zJ_x-J_yJ_zJ_y$, $J_xJ_zJ_y+J_yJ_zJ_x)$\\
%${\rm C_6}$ & A & $\bm{k} \cdot \bm{J}$ & $C_2$ &A &$J_z$\\
% &  & & &(${\rm E_2}$) &$([J_xJ_zJ_x-J_yJ_zJ_y, J_xJ_zJ_y+J_yJ_zJ_x])$\\
 ${\rm C_s}$ & A & $k_x J_z, k_xJ_y, k_y J_x , k_zJ_x$&$\sigma_v (yz)$ &A& $J_x$\\
${\rm C_{2v}}$ & ${\rm A_2}$ & $k_z J_z$&$C_2$&${\rm A_2}$&$J_z$\\
% &  & &&(${\rm A_1}$) & $(J_xJ_zJ_y, J_yJ_zJ_x)$\\
 & ${\rm B_1}$ &$k_x J_z, k_z J_x$ &$\sigma_v (yz)$&${\rm B_1}$ & $J_x$\\
%  & & &&(${\rm A_1}$)& $(J_xJ_zJ_y, J_yJ_zJ_x)$\\
 & ${\rm B_2}$ &$k_yJ_z ,  k_z J_y$ &$\sigma_v(xz)$&${\rm B_2}$ & $J_y$\\
% & & &&(${\rm A_1}$)& $(J_xJ_zJ_y, J_yJ_zJ_x)$\\
${\rm C_{3v}}$ & ${\rm A_1}$&$k_z (J_x^3-J_xJ_yJ_y-J_yJ_xJ_y-J_yJ_yJ_x)$ &$\sigma_v (yz)$&${\rm A_1}$&$J_x^3-J_xJ_yJ_y-J_yJ_xJ_y-J_yJ_yJ_x$\\
${\rm C_{4v}}$ & ${\rm A_2}$&$k_z J_z$ &$C_2$&${\rm A_2}$&$J_z$\\
% & & &&(${\rm A_1}$, ${\rm B_1}$, ${\rm B_2}$)&($-$, $J_xJ_zJ_y+J_yJ_zJ_x$, $J_xJ_zJ_x-J_yJ_zJ_y$)\\
${\rm C_{6v}}$ & ${\rm A_2}$&$k_z J_z$ &$C_2$&${\rm A_2}$&$J_z$\\
% & & &&(${\rm A_1}$, ${\rm E_2}$) &($-$, $[J_xJ_zJ_x-J_yJ_zJ_y, J_xJ_zJ_y+J_yJ_zJ_x]$)\\ 
& ${\rm B_1}$& $k_z(J_x^3-J_xJ_yJ_y-J_yJ_xJ_y-J_yJ_yJ_x)$&$\sigma_v(yz)$&${\rm B_1}$&$J_x^3-J_xJ_yJ_y-J_yJ_xJ_y-J_yJ_yJ_x$\\
 & ${\rm B_2}$& $k_z(J_y^3-J_yJ_xJ_x-J_xJ_yJ_x-J_xJ_xJ_y)$&$\sigma_d(xz)$&${\rm B_2}$&$J_y^3-J_yJ_xJ_x-J_xJ_yJ_x-J_xJ_xJ_y$\\
\hline\hline
\end{tabular} 
\end{table*}

To systematically study the magnetic response of MFs, we examine possible contributions of MFs to local operators
$
\hat{O}(x)=\hat{c}_{\sigma}^{\dagger}(x)O_{\sigma, \sigma'}\hat{c}_{\sigma'}(x) 
$
of electrons, where $\hat{c}_{\sigma}^{\dagger}(x)$ and $\hat{c}_{\sigma}(x)$ are the electron operators and $\sigma$ is the internal degrees of freedom such as spin, orbital, and so on.
To obtain a physical response, the matrix $O_{\sigma, \sigma'}$ should be Hermitian. 
The MFs have a nonzero response to external fields 
through local operators $\hat{O}(x)$. 
For instance, if MFs make a nonzero contribution to the electron-spin operator $\hat{S}_i(x)=\hat{c}_{\sigma}^{\dagger}(x)[s_i/2]_{\sigma, \sigma'}\hat{c}_{\sigma'}(x)$ with a Pauli matrix $s_i$, then the MF shows a nonzero magnetic response through the Zeeman term of electrons. 

In the Nambu space with $\hat{\Psi}^{\dagger}(x)=(\hat{c}^{\dagger}_{\sigma}(x), \hat{c}_{\sigma}(x))$, $\hat{O}(x)$ is recast into
$
\hat{O}(x)=(1/2)\hat{\Psi}^{\dagger}(x){\cal O}\hat{\Psi}(x),
$
with ${\cal O}={\rm diag}(O,-O^T)={\rm diag}(O, -O^*)$, where we have used the Hermiticity of $O$.
Next, by expanding the mode of the quantum field
$
\hat{\Psi}(x)=\sum_a \hat{\gamma}^{(a)}|u^{(a)}_0\rangle + (\text{nonzero modes}), 
$
we obtain the coupling between $\hat{O}(x)$ and the MFs $\hat{\gamma}^{(a)}$ in the low-energy limit,
\begin{align}
\hat{O}_{\rm MF}&=\frac{1}{2}\sum_{ab}\hat{\gamma}^{(a)}\hat{\gamma}^{(b)}
\langle u_0^{(a)}|{\cal O}|u_0^{(b)}\rangle
\nonumber\\
&=-\frac{i}{4}\sum_{ab}\hat{\gamma}^{(a)}\hat{\gamma}^{(b)}{\rm tr}
\left[{\cal O}
\rho^{(ab)} 
\right],
\label{eq:OMF}
\end{align}
where $\rho^{(ab)} \equiv i\left(|u_0^{(b)}\rangle\langle u_0^{(a)}|-|u_0^{(a)}\rangle\langle u_0^{(b)}|
\right)$. In this formalism, crystalline symmetry is properly considered by the irreducible decomposition of ${\cal O}$ as 
$
{\cal O}=\sum_{\Gamma}{\cal O}_{\Gamma}, 
$
where ${\cal O}_{\Gamma}$ is an IR of the point group on the surface.
As shown below, $|u_0^{(a)}\rangle$ obeys several symmetry constraints, so only a few representations of ${\cal O}_{\Gamma}$ can provide nonzero contributions in Eq.~(\ref{eq:OMF}).

We now discuss the symmetry constraints.
First, to have a nonzero $w_{\rm M1D}$, we need a bulk superconducting gap at the high-symmetry line or plane on which $w_{\rm M1D}$ is defined. 
This requirement restricts the possible pairing symmetry of Cooper pairs.
The pairing symmetry must also maintain TR symmetry because we consider 3D TR-invariant TSCs.
Moreover, if the bulk system has inversion symmetry, the pairing symmetry must be odd under inversion~\cite{Xiong17,suppl}. 
Second, the zero modes $|u_0^{(a)}\rangle$ should be a representation of the point group that is compatible with both the surface and the pairing symmetries because the BdG Hamiltonian respects these symmetries.
Note that nonconventional Cooper pairs spontaneously break part of the crystalline symmetry, so
the zero modes respect only the unbroken part. 
Third, a much stronger symmetry constraint is obtained from the index theorem of $w_{\rm M1D}$~\cite{Sato11}.
The index theorem says that 
zero modes $|u^{(a)}_0\rangle$ 
should be eigenstates of $\Gamma_{\rm M}$; for example,
\begin{align}
\Gamma_{\rm M}|u_0^{(a)}\rangle=|u_0^{(a)}\rangle.
\label{eq:chirality}
\end{align}
Here all stable zero modes should have the same eigenvalue $\Gamma_{\rm M}$, otherwise zero modes with opposite eigenvalues are easily gapped in pairs, even by a symmetry-preserving perturbation. 
In fact, this important property can be rigorously proven for generic lattice systems~\cite{Xiong17}. 
Finally, for the zero modes to exist, any surface-preserving point-group operation for the BdG Hamiltonian should not anticommute with $\Gamma_{\rm M}$.
The last claim is proven by contradiction.
If such a point group operation exists, one can generate another zero mode whose eigenvalue is the opposite of $\Gamma_{\rm M}$ by operating on a zero mode with the point group. This contradicts the above property of the zero modes, so the claim holds.

Note that the last constraint also restricts any possible pairing symmetry of Cooper pairs. 
In a nonconventional superconductor, depending on the pairing symmetry, crystalline symmetry can be realized projectively as a combination with a $U(1)$ gauge rotation, which changes their commutation to that of the chiral operator $\Gamma_{\rm M}$.
For instance, if the gap function (Cooper pair) is odd under a mirror reflection, then the mirror reflection of the BdG Hamiltonian is the original reflection combined with a $U(1)$ $\pi$ rotation, so it anticommutes with the PH operator. Therefore, its commutation with $\Gamma_{\rm M}$ changes.
As discussed above, stable zero modes only exist when any surface-preserving point-group operation does not anticommute with $\Gamma_{\rm M}$, which restricts any possible pairing symmetry between Cooper pairs~\cite{suppl}.

Based on these arguments, we determine the possible pairing symmetry of Cooper pairs and IRs of $\rho^{(ab)}$~\cite{suppl}.
We also find that only the same IRs of ${\cal O}_{\Gamma}$ give nonzero contributions in Eq. (\ref{eq:OMF}).
Table \ref{table} summarizes the IRs of ${\cal O}_{\Gamma}$ and pairing symmetry that satisfy the above constraints. 
Remarkably, the results show that the IRs of $\Delta_{\Gamma}$ and those of $\mathcal{O}_{\Gamma}$ coincide with each other up to leading order. 
This notable property allows us to determine the pairing symmetry through the magnetic response of MFs.
The results also show that a magnetic-octupole response is possible when the surface has $C_{\rm 3v}$ or $C_{\rm 6v}$ symmetry. 
As shown below, the octupole response only appears for MFs in high-spin TSCs of spin-3/2 electrons.
The order of magnetic multipoles reflects a difference between TSCs with spin $1/2$ and $3/2$.

\begin{figure}[tbp]
\centering
 \includegraphics[width=8cm]{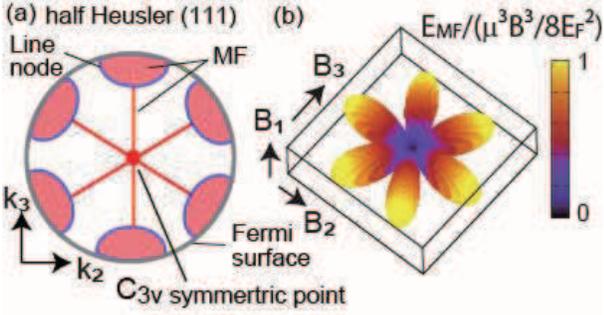}
\caption{ (a) Surface states of half-Heusler superconductor in ($111$) plane. The red line and red areas indicate helical MFs with flat dispersion and the line-node-induced Majorana flat bands. $k_1 = \frac{1}{\sqrt{3}} (k_x+k_y+k_z)$, $k_2=\frac{1}{\sqrt{2}}(k_x-k_y)$, and $k_3= \frac{1}{\sqrt{6}}(k_x+k_y-2k_z)$. (b) Energy gap of helical MF at $k_2=k_3=0$ as a function of $\bm{B}$ under the Zeeman magnetic field $\mu \bm{B} \cdot \bm{J}$.} \label{fig:multipole}
\end{figure}

\paragraph*{Majorana octupole in spin-3/2 superconductors.}
The results presented in Table \ref{table} indicate that helical MFs on a surface-preserving $C_{3v}$ or $C_{6v}$ host the magnetic octupole. This unique behavior is intrinsic to high-spin TSCs of spin-3/2 electrons for the following reasons.

First, the base of $O_{\Gamma}$ for the magnetic octupole vanishes if the $J_i$ are given by the Pauli matrices of spin-1/2 electrons. In fact, we have $J_x^3-J_xJ_yJ_y-J_yJ_xJ_y-J_yJ_yJ_x=J_y^3-J_yJ_xJ_x-J_xJ_yJ_x-J_xJ_xJ_y=0$ for $J_i=\sigma_i/2$.
Furthermore, if the pairing symmetry is $A_1$ ($B_{i=1,2}$) for $C_{\rm 3v}$ ($C_{\rm 6v}$), which is required for the octupole response, 
the spin-1/2 superconductor hosts a superconducting node at a high-symmetry line, so 
it cannot support well-defined helical MFs with magnetic octupoles because 
$C_3$ symmetry for spin-1/2 electrons is enhanced to $C_{\infty}$ on the axis of rotation in the Brillouin zone ~\cite{suppl}.

Contrastingly, for spin-3/2 superconductors, helical MFs exhibit an octupole response.
To illustrate this, we calculate the magnetic response of MFs in half-Heusler compounds.
In these compounds~\cite{Goll08,Butch11,Tafti13,GXu16,Bay12,Kim18}, a strong spin-orbit interaction (SOI) and high crystal symmetry provide a fourfold degenerate band at the $\Gamma$ point, which is well described by spin-$3/2$ fermions~\cite{Brydon16}. Additionally, recent experiments have suggested the existence of parity-mixed superconductivity with line nodes~\cite{Bay12,Kim18}. We show here that the parity-mixed superconductor exhibits a magnetic-octupole response. Consider the low-energy model with $T_d$ symmetry~\cite{Brydon16}:
\begin{align}
H_{\rm LK}(\bm{k}) =& \alpha \bm{k}^2 + \beta \sum_{i} k_i J_i^2 + \gamma \sum_{i \neq j} k_i k_j J_i J_j \notag \\
& + \delta \sum_{i}k_i(J_{i+1}J_i J_{i+1}-J_{i+2}J_{i}J_{i+2}), \label{eq:LK-model}
\end{align}
where $i=x,y,z$ and $i+1=y$ if $i=x$, etc., and $J_i$ are the $4 \times 4$ spin matrices of spin-$3/2$ fermions. Because inversion symmetry is absent, the Hamiltonian includes the antisymmetric SOI, which is proportional to $\delta$ and causes spin splitting at the Fermis surface~\cite{Kim18}. In their superconducting states, Cooper pairs form between spin-$3/2$ electrons, which allows quintet and septet parings in addition to the conventional singlet and triplet pairings~\cite{Ho99,WYang16,Brydon16}. Furthermore, the antisymmetric SOI generally mixes the parity of the gap function, so the even- and odd-parity components coexist in the gap function~\cite{Gorkov01,Frigeri04,Fujimoto07a,Fujimoto07b,Bauer12} and the odd-parity component is aligned with the antisymmetric SOI~\cite{Frigeri04}, providing the spin-septet pairing~\cite{Brydon16,Kim18}. Based on this insight, the gap function must include the spin-septet component, $\Delta(\bm{k}) = \Delta/{\sqrt{1+\eta^2}}[ \eta 1_4 + \sum_{i} k_i (J_{i+1}J_i J_{i+1}-J_{i+2}J_{i}J_{i+2})](e^{-i J_y \pi})$, in addition to an $s$-wave singlet state, even when we choose the conventional $A_1$ state of $T_d$, where $\eta$ parametrizes the mixing between the $s$-wave and spin-septet components and $1_n$ is the $n \times n$ identity matrix. Here, the PH, TR, and $T_d$ symmetry operations hosted by the BdG Hamiltonian are $\mathcal{C} = \tau_x K$, $\mathcal{T}=e^{-i J_y \pi } K$, and $\diag [e^{-i\frac{i 2\pi}{q} \bm{J} \cdot \bm{n}},e^{i\frac{i 2\pi}{q} \bm{J} \cdot \bm{n}}]$, respectively. 

The superconducting state hosts six line nodes encircling the $k_x$, $k_y$, and $k_z$ axis~\cite{Brydon16}, in analogy with other parity-mixed superconductors~\cite{Tanaka10,Yada11,Sato11,Brydon11,Schnyder12,Matsuura13,Schnyder15}. Here, we focus on the (111) surface because the magnetic-octupole response requires $C_{\rm 3v}$ symmetry. 
To verify the existence of helical MFs, we numerically diagonalize the BdG Hamiltonian with the surface normal to the $[111]$ direction and find a helical MF with three flat dispersion curves (see Fig. S2 in the Supplemental Material \cite{suppl}), as schematically depicted in Fig.~\ref{fig:multipole}(c). Each flat dispersion curve lies on the mirror planes with mirror-reflection symmetries, $\sigma$, $C_3^{\dagger} \sigma C_3$, and $(C_3^{\dagger})^2 \sigma (C_3)^2$, where $\sigma$ is mirror-reflection with respect to the $(1\bar{1}0)$ plane and $C_3$ is a threefold rotation around the $[111]$ direction. Combining these mirror reflections with PH and TR operations, we obtain three $\Gamma_{\rm M}$ and the associated $w_{\rm M1D}$, which protects zero modes on each flat dispersion curve. In particular, the three flat dispersion curves meet at a $C_{\rm 3v}$ symmetry point. 

Based on the constraint~(\ref{eq:chirality}), the zero modes can be simultaneous eigenstates of $\Gamma_{\rm M}$ and of $C_3$. In this case, we have $\Gamma_{\rm M} | u_0^{(a)} \rangle =| u_0^{(a)} \rangle$ and $C_3 | u_0^{(a)} \rangle =-| u_0^{(a)} \rangle$ with $a=1,2$ being the label for a Kramers pair~\cite{suppl}, which lead to $\Gamma_{\rm M} \rho^{(12)} \Gamma_{\rm M}^{-1} =C_3 \rho^{(12)} C_3^{-1}=\rho^{(12)}$. Thus, $\mathcal{O}_{\Gamma}$ needs to be the trivial representation $A_1$ in $C_{\rm 3v}$, as shown in Table~\ref{table}. 
To demonstrate magnetic response, we add a Zeeman magnetic term $\mu \bm{B} \cdot \bm{J}$ in Eq.~(\ref{eq:LK-model}), which leads to an anisotropic response with $C_3$ symmetry in Fig.~\ref{fig:multipole}(d). The Zeeman magnetic term contributes to the energy gap of the MFs on the order of $3 \sqrt{2} \mu^3 B^3/32E_F^2$, where $E_F$ is the Fermi energy~\cite{suppl}, implying a magnetic-octupole response.

Another high-spin superconductor of spin-3/2 electrons was recently proposed for antiperovskite materials with $O_h$ group~\cite{Oudah16,Kawakami18}.
We obtain a similar magnetic-octupole response of MFs on the $(111)$ surface when its pairing symmetry is $A_{2u}$ of $O_h$.

For comparison, we also examine magnetic response of helical MFs in the doped superconducting TI, {\it A}$_{x}$Bi$_2$Se$_3$ ($A$=Cu, Sr, Nb),
which becomes a TSC when an odd-parity Cooper pair is realized~\cite{Sato09,Sato10,Fu10,Fu14, Matano16, Yonezawa16}. 
Since Bi$_2$Se$_3$ has $D_{3d}$ symmetry, the surface normal to the $c$ axis (i.e., (111) surface) hosts $C_{\rm 3v}$ symmetry like the half-Heusler case. However, the doped TI merely exhibits the magnetic-dipole response of MFs to leading order, or it cannot host a well-define helical MFs on the (111) surface~\cite{suppl,Sumita18}, since it is a conventional spin-1/2 TSC.

\paragraph*{Conclusions.} 
In this paper, we develop a theory of Majorana multipoles for 3D TR-invariant TSCs, which provide novel experimental means to identify bulk pairing symmetry and high-spin superconductivity. 
The Majorana multipoles may be observed through spin-sensitive measurements such as spatially resolved NMR measurements~\cite{Chung09} or the surface tunneling spectroscopy under magnetic fields~\cite{Fogelstrom97,Tanaka95,Tanaka02,Tanuma02,Tanaka09,Tamura17}. 

This work was supported by the Grants-in-Aid for Scientific Research on Innovative Areas ``Topological Material Science'' (Grant Nos. JP15H05855, JP15H05851, JP15H05853, JP15K21717, and JP18H04224) from JSPS of Japan, and by JSPS KAKENHI Grant Nos. JP17H02922 and JP18H01176. S.K. was supported by the CREST project (JPMJCR16F2) from Japan Science and Technology Agency (JST), and the Building of Consortia for the Development of Human Resources in Science and Technology.

\bibliography{multipole}

\begin{thebibliography}{79}%
\makeatletter
\providecommand \@ifxundefined [1]{%
 \@ifx{#1\undefined}
}%
\providecommand \@ifnum [1]{%
 \ifnum #1\expandafter \@firstoftwo
 \else \expandafter \@secondoftwo
 \fi
}%
\providecommand \@ifx [1]{%
 \ifx #1\expandafter \@firstoftwo
 \else \expandafter \@secondoftwo
 \fi
}%
\providecommand \natexlab [1]{#1}%
\providecommand \enquote  [1]{``#1''}%
\providecommand \bibnamefont  [1]{#1}%
\providecommand \bibfnamefont [1]{#1}%
\providecommand \citenamefont [1]{#1}%
\providecommand \href@noop [0]{\@secondoftwo}%
\providecommand \href [0]{\begingroup \@sanitize@url \@href}%
\providecommand \@href[1]{\@@startlink{#1}\@@href}%
\providecommand \@@href[1]{\endgroup#1\@@endlink}%
\providecommand \@sanitize@url [0]{\catcode `\\12\catcode `\$12\catcode
  `\&12\catcode `\#12\catcode `\^12\catcode `\_12\catcode `\%12\relax}%
\providecommand \@@startlink[1]{}%
\providecommand \@@endlink[0]{}%
\providecommand \url  [0]{\begingroup\@sanitize@url \@url }%
\providecommand \@url [1]{\endgroup\@href {#1}{\urlprefix }}%
\providecommand \urlprefix  [0]{URL }%
\providecommand \Eprint [0]{\href }%
\providecommand \doibase [0]{http://dx.doi.org/}%
\providecommand \selectlanguage [0]{\@gobble}%
\providecommand \bibinfo  [0]{\@secondoftwo}%
\providecommand \bibfield  [0]{\@secondoftwo}%
\providecommand \translation [1]{[#1]}%
\providecommand \BibitemOpen [0]{}%
\providecommand \bibitemStop [0]{}%
\providecommand \bibitemNoStop [0]{.\EOS\space}%
\providecommand \EOS [0]{\spacefactor3000\relax}%
\providecommand \BibitemShut  [1]{\csname bibitem#1\endcsname}%
\let\auto@bib@innerbib\@empty
%</preamble>
\bibitem [{\citenamefont {Hu}(1994)}]{Hu94}%
  \BibitemOpen
  \bibfield  {author} {\bibinfo {author} {\bibfnamefont {C.-R.}\ \bibnamefont
  {Hu}},\ }\href {\doibase 10.1103/PhysRevLett.72.1526} {\bibfield  {journal}
  {\bibinfo  {journal} {Phys. Rev. Lett.}\ }\textbf {\bibinfo {volume} {72}},\
  \bibinfo {pages} {1526} (\bibinfo {year} {1994})}\BibitemShut {NoStop}%
\bibitem [{\citenamefont {Kashiwaya}\ and\ \citenamefont
  {Tanaka}(2000)}]{Kashiwaya2000}%
  \BibitemOpen
  \bibfield  {author} {\bibinfo {author} {\bibfnamefont {S.}~\bibnamefont
  {Kashiwaya}}\ and\ \bibinfo {author} {\bibfnamefont {Y.}~\bibnamefont
  {Tanaka}},\ }\href {https://doi.org/10.1088/0034-4885/63/10/202} {\bibfield
  {journal} {\bibinfo  {journal} {Rep. Prog. Phys.}\ }\textbf {\bibinfo
  {volume} {63}},\ \bibinfo {pages} {1641} (\bibinfo {year}
  {2000})}\BibitemShut {NoStop}%
\bibitem [{\citenamefont {Volovik}(2003)}]{Volovik03}%
  \BibitemOpen
  \bibfield  {author} {\bibinfo {author} {\bibfnamefont {G.~E.}\ \bibnamefont
  {Volovik}},\ }\href@noop {} {\emph {\bibinfo {title} {The Universe in a
  Helium Droplet}}}\ (\bibinfo  {publisher} {Oxford University Press},\
  \bibinfo {address} {Oxford},\ \bibinfo {year} {2003})\BibitemShut {NoStop}%
\bibitem [{\citenamefont {Schnyder}\ \emph {et~al.}(2008)\citenamefont
  {Schnyder}, \citenamefont {Ryu}, \citenamefont {Furusaki},\ and\
  \citenamefont {Ludwig}}]{Schnyder08}%
  \BibitemOpen
  \bibfield  {author} {\bibinfo {author} {\bibfnamefont {A.~P.}\ \bibnamefont
  {Schnyder}}, \bibinfo {author} {\bibfnamefont {S.}~\bibnamefont {Ryu}},
  \bibinfo {author} {\bibfnamefont {A.}~\bibnamefont {Furusaki}}, \ and\
  \bibinfo {author} {\bibfnamefont {A.~W.~W.}\ \bibnamefont {Ludwig}},\ }\href
  {\doibase 10.1103/PhysRevB.78.195125} {\bibfield  {journal} {\bibinfo
  {journal} {Phys. Rev. B}\ }\textbf {\bibinfo {volume} {78}},\ \bibinfo
  {pages} {195125} (\bibinfo {year} {2008})}\BibitemShut {NoStop}%
\bibitem [{\citenamefont {Sato}(2009)}]{Sato09}%
  \BibitemOpen
  \bibfield  {author} {\bibinfo {author} {\bibfnamefont {M.}~\bibnamefont
  {Sato}},\ }\href {\doibase 10.1103/PhysRevB.79.214526} {\bibfield  {journal}
  {\bibinfo  {journal} {Phys. Rev. B}\ }\textbf {\bibinfo {volume} {79}},\
  \bibinfo {pages} {214526} (\bibinfo {year} {2009})}\BibitemShut {NoStop}%
\bibitem [{\citenamefont {Wilczek}(2009)}]{Wilczek09}%
  \BibitemOpen
  \bibfield  {author} {\bibinfo {author} {\bibfnamefont {F.}~\bibnamefont
  {Wilczek}},\ }\href {http://dx.doi.org/10.1038/nphys1380} {\bibfield
  {journal} {\bibinfo  {journal} {Nature Phys.}\ }\textbf {\bibinfo {volume}
  {5}},\ \bibinfo {pages} {614} (\bibinfo {year} {2009})}\BibitemShut {NoStop}%
\bibitem [{\citenamefont {Hasan}\ and\ \citenamefont {Kane}(2010)}]{Hasan10}%
  \BibitemOpen
  \bibfield  {author} {\bibinfo {author} {\bibfnamefont {M.~Z.}\ \bibnamefont
  {Hasan}}\ and\ \bibinfo {author} {\bibfnamefont {C.~L.}\ \bibnamefont
  {Kane}},\ }\href {\doibase 10.1103/RevModPhys.82.3045} {\bibfield  {journal}
  {\bibinfo  {journal} {Rev. Mod. Phys.}\ }\textbf {\bibinfo {volume} {82}},\
  \bibinfo {pages} {3045} (\bibinfo {year} {2010})}\BibitemShut {NoStop}%
\bibitem [{\citenamefont {Qi}\ and\ \citenamefont {Zhang}(2011)}]{Qi11}%
  \BibitemOpen
  \bibfield  {author} {\bibinfo {author} {\bibfnamefont {X.-L.}\ \bibnamefont
  {Qi}}\ and\ \bibinfo {author} {\bibfnamefont {S.-C.}\ \bibnamefont {Zhang}},\
  }\href {\doibase 10.1103/RevModPhys.83.1057} {\bibfield  {journal} {\bibinfo
  {journal} {Rev. Mod. Phys.}\ }\textbf {\bibinfo {volume} {83}},\ \bibinfo
  {pages} {1057} (\bibinfo {year} {2011})}\BibitemShut {NoStop}%
\bibitem [{\citenamefont {Tanaka}\ \emph {et~al.}(2012)\citenamefont {Tanaka},
  \citenamefont {Sato},\ and\ \citenamefont {Nagaosa}}]{Tanaka12}%
  \BibitemOpen
  \bibfield  {author} {\bibinfo {author} {\bibfnamefont {Y.}~\bibnamefont
  {Tanaka}}, \bibinfo {author} {\bibfnamefont {M.}~\bibnamefont {Sato}}, \ and\
  \bibinfo {author} {\bibfnamefont {N.}~\bibnamefont {Nagaosa}},\ }\href
  {\doibase 10.1143/JPSJ.81.011013} {\bibfield  {journal} {\bibinfo  {journal}
  {Journal of the Physical Society of Japan}\ }\textbf {\bibinfo {volume}
  {81}},\ \bibinfo {pages} {011013} (\bibinfo {year} {2012})}\BibitemShut
  {NoStop}%
\bibitem [{\citenamefont {Alicea}(2012)}]{Alicea12}%
  \BibitemOpen
  \bibfield  {author} {\bibinfo {author} {\bibfnamefont {J.}~\bibnamefont
  {Alicea}},\ }\href
  {http://iopscience.iop.org/article/10.1088/0034-4885/75/7/076501} {\bibfield
  {journal} {\bibinfo  {journal} {Rep. Prog. Phys.}\ }\textbf {\bibinfo
  {volume} {75}},\ \bibinfo {pages} {076501} (\bibinfo {year}
  {2012})}\BibitemShut {NoStop}%
\bibitem [{\citenamefont {Sato}\ and\ \citenamefont
  {Fujimoto}(2016)}]{SatoFujimoto16}%
  \BibitemOpen
  \bibfield  {author} {\bibinfo {author} {\bibfnamefont {M.}~\bibnamefont
  {Sato}}\ and\ \bibinfo {author} {\bibfnamefont {S.}~\bibnamefont
  {Fujimoto}},\ }\href {\doibase 10.7566/JPSJ.85.072001} {\bibfield  {journal}
  {\bibinfo  {journal} {Journal of the Physical Society of Japan}\ }\textbf
  {\bibinfo {volume} {85}},\ \bibinfo {pages} {072001} (\bibinfo {year}
  {2016})}\BibitemShut {NoStop}%
\bibitem [{\citenamefont {Mizushima}\ \emph {et~al.}(2016)\citenamefont
  {Mizushima}, \citenamefont {Tsutsumi}, \citenamefont {Kawakami},
  \citenamefont {Sato}, \citenamefont {Ichioka},\ and\ \citenamefont
  {Machida}}]{Mizushima16}%
  \BibitemOpen
  \bibfield  {author} {\bibinfo {author} {\bibfnamefont {T.}~\bibnamefont
  {Mizushima}}, \bibinfo {author} {\bibfnamefont {Y.}~\bibnamefont {Tsutsumi}},
  \bibinfo {author} {\bibfnamefont {T.}~\bibnamefont {Kawakami}}, \bibinfo
  {author} {\bibfnamefont {M.}~\bibnamefont {Sato}}, \bibinfo {author}
  {\bibfnamefont {M.}~\bibnamefont {Ichioka}}, \ and\ \bibinfo {author}
  {\bibfnamefont {K.}~\bibnamefont {Machida}},\ }\href {\doibase
  10.7566/JPSJ.85.022001} {\bibfield  {journal} {\bibinfo  {journal} {Journal
  of the Physical Society of Japan}\ }\textbf {\bibinfo {volume} {85}},\
  \bibinfo {pages} {022001} (\bibinfo {year} {2016})}\BibitemShut {NoStop}%
\bibitem [{\citenamefont {Chiu}\ \emph {et~al.}(2016)\citenamefont {Chiu},
  \citenamefont {Teo}, \citenamefont {Schnyder},\ and\ \citenamefont
  {Ryu}}]{Chiu16}%
  \BibitemOpen
  \bibfield  {author} {\bibinfo {author} {\bibfnamefont {C.-K.}\ \bibnamefont
  {Chiu}}, \bibinfo {author} {\bibfnamefont {J.~C.~Y.}\ \bibnamefont {Teo}},
  \bibinfo {author} {\bibfnamefont {A.~P.}\ \bibnamefont {Schnyder}}, \ and\
  \bibinfo {author} {\bibfnamefont {S.}~\bibnamefont {Ryu}},\ }\href {\doibase
  10.1103/RevModPhys.88.035005} {\bibfield  {journal} {\bibinfo  {journal}
  {Rev. Mod. Phys.}\ }\textbf {\bibinfo {volume} {88}},\ \bibinfo {pages}
  {035005} (\bibinfo {year} {2016})}\BibitemShut {NoStop}%
\bibitem [{\citenamefont {Sato}\ and\ \citenamefont {Ando}(2017)}]{Sato17}%
  \BibitemOpen
  \bibfield  {author} {\bibinfo {author} {\bibfnamefont {M.}~\bibnamefont
  {Sato}}\ and\ \bibinfo {author} {\bibfnamefont {Y.}~\bibnamefont {Ando}},\
  }\href {http://stacks.iop.org/0034-4885/80/i=7/a=076501} {\bibfield
  {journal} {\bibinfo  {journal} {Rep. Prog. Phys.}\ }\textbf {\bibinfo
  {volume} {80}},\ \bibinfo {pages} {076501} (\bibinfo {year}
  {2017})}\BibitemShut {NoStop}%
\bibitem [{\citenamefont {Nayak}\ \emph {et~al.}(2008)\citenamefont {Nayak},
  \citenamefont {Simon}, \citenamefont {Stern}, \citenamefont {Freedman},\ and\
  \citenamefont {Das~Sarma}}]{Nayak08}%
  \BibitemOpen
  \bibfield  {author} {\bibinfo {author} {\bibfnamefont {C.}~\bibnamefont
  {Nayak}}, \bibinfo {author} {\bibfnamefont {S.~H.}\ \bibnamefont {Simon}},
  \bibinfo {author} {\bibfnamefont {A.}~\bibnamefont {Stern}}, \bibinfo
  {author} {\bibfnamefont {M.}~\bibnamefont {Freedman}}, \ and\ \bibinfo
  {author} {\bibfnamefont {S.}~\bibnamefont {Das~Sarma}},\ }\href {\doibase
  10.1103/RevModPhys.80.1083} {\bibfield  {journal} {\bibinfo  {journal} {Rev.
  Mod. Phys.}\ }\textbf {\bibinfo {volume} {80}},\ \bibinfo {pages} {1083}
  (\bibinfo {year} {2008})}\BibitemShut {NoStop}%
\bibitem [{\citenamefont {Hor}\ \emph {et~al.}(2010)\citenamefont {Hor},
  \citenamefont {Williams}, \citenamefont {Checkelsky}, \citenamefont
  {Roushan}, \citenamefont {Seo}, \citenamefont {Xu}, \citenamefont
  {Zandbergen}, \citenamefont {Yazdani}, \citenamefont {Ong},\ and\
  \citenamefont {Cava}}]{Hor10}%
  \BibitemOpen
  \bibfield  {author} {\bibinfo {author} {\bibfnamefont {Y.~S.}\ \bibnamefont
  {Hor}}, \bibinfo {author} {\bibfnamefont {A.~J.}\ \bibnamefont {Williams}},
  \bibinfo {author} {\bibfnamefont {J.~G.}\ \bibnamefont {Checkelsky}},
  \bibinfo {author} {\bibfnamefont {P.}~\bibnamefont {Roushan}}, \bibinfo
  {author} {\bibfnamefont {J.}~\bibnamefont {Seo}}, \bibinfo {author}
  {\bibfnamefont {Q.}~\bibnamefont {Xu}}, \bibinfo {author} {\bibfnamefont
  {H.~W.}\ \bibnamefont {Zandbergen}}, \bibinfo {author} {\bibfnamefont
  {A.}~\bibnamefont {Yazdani}}, \bibinfo {author} {\bibfnamefont {N.~P.}\
  \bibnamefont {Ong}}, \ and\ \bibinfo {author} {\bibfnamefont {R.~J.}\
  \bibnamefont {Cava}},\ }\href {\doibase 10.1103/PhysRevLett.104.057001}
  {\bibfield  {journal} {\bibinfo  {journal} {Phys. Rev. Lett.}\ }\textbf
  {\bibinfo {volume} {104}},\ \bibinfo {pages} {057001} (\bibinfo {year}
  {2010})}\BibitemShut {NoStop}%
\bibitem [{\citenamefont {Fu}\ and\ \citenamefont {Berg}(2010)}]{Fu10}%
  \BibitemOpen
  \bibfield  {author} {\bibinfo {author} {\bibfnamefont {L.}~\bibnamefont
  {Fu}}\ and\ \bibinfo {author} {\bibfnamefont {E.}~\bibnamefont {Berg}},\
  }\href {\doibase 10.1103/PhysRevLett.105.097001} {\bibfield  {journal}
  {\bibinfo  {journal} {Phys. Rev. Lett.}\ }\textbf {\bibinfo {volume} {105}},\
  \bibinfo {pages} {097001} (\bibinfo {year} {2010})}\BibitemShut {NoStop}%
\bibitem [{\citenamefont {Sasaki}\ \emph {et~al.}(2011)\citenamefont {Sasaki},
  \citenamefont {Kriener}, \citenamefont {Segawa}, \citenamefont {Yada},
  \citenamefont {Tanaka}, \citenamefont {Sato},\ and\ \citenamefont
  {Ando}}]{Sasaki11}%
  \BibitemOpen
  \bibfield  {author} {\bibinfo {author} {\bibfnamefont {S.}~\bibnamefont
  {Sasaki}}, \bibinfo {author} {\bibfnamefont {M.}~\bibnamefont {Kriener}},
  \bibinfo {author} {\bibfnamefont {K.}~\bibnamefont {Segawa}}, \bibinfo
  {author} {\bibfnamefont {K.}~\bibnamefont {Yada}}, \bibinfo {author}
  {\bibfnamefont {Y.}~\bibnamefont {Tanaka}}, \bibinfo {author} {\bibfnamefont
  {M.}~\bibnamefont {Sato}}, \ and\ \bibinfo {author} {\bibfnamefont
  {Y.}~\bibnamefont {Ando}},\ }\href {\doibase 10.1103/PhysRevLett.107.217001}
  {\bibfield  {journal} {\bibinfo  {journal} {Phys. Rev. Lett.}\ }\textbf
  {\bibinfo {volume} {107}},\ \bibinfo {pages} {217001} (\bibinfo {year}
  {2011})}\BibitemShut {NoStop}%
\bibitem [{\citenamefont {Sasaki}\ \emph {et~al.}(2012)\citenamefont {Sasaki},
  \citenamefont {Ren}, \citenamefont {Taskin}, \citenamefont {Segawa},
  \citenamefont {Fu},\ and\ \citenamefont {Ando}}]{Sasaki12}%
  \BibitemOpen
  \bibfield  {author} {\bibinfo {author} {\bibfnamefont {S.}~\bibnamefont
  {Sasaki}}, \bibinfo {author} {\bibfnamefont {Z.}~\bibnamefont {Ren}},
  \bibinfo {author} {\bibfnamefont {A.~A.}\ \bibnamefont {Taskin}}, \bibinfo
  {author} {\bibfnamefont {K.}~\bibnamefont {Segawa}}, \bibinfo {author}
  {\bibfnamefont {L.}~\bibnamefont {Fu}}, \ and\ \bibinfo {author}
  {\bibfnamefont {Y.}~\bibnamefont {Ando}},\ }\href {\doibase
  10.1103/PhysRevLett.109.217004} {\bibfield  {journal} {\bibinfo  {journal}
  {Phys. Rev. Lett.}\ }\textbf {\bibinfo {volume} {109}},\ \bibinfo {pages}
  {217004} (\bibinfo {year} {2012})}\BibitemShut {NoStop}%
\bibitem [{\citenamefont {Hashimoto}\ \emph {et~al.}(2015)\citenamefont
  {Hashimoto}, \citenamefont {Yada}, \citenamefont {Sato},\ and\ \citenamefont
  {Tanaka}}]{Hashimoto15}%
  \BibitemOpen
  \bibfield  {author} {\bibinfo {author} {\bibfnamefont {T.}~\bibnamefont
  {Hashimoto}}, \bibinfo {author} {\bibfnamefont {K.}~\bibnamefont {Yada}},
  \bibinfo {author} {\bibfnamefont {M.}~\bibnamefont {Sato}}, \ and\ \bibinfo
  {author} {\bibfnamefont {Y.}~\bibnamefont {Tanaka}},\ }\href {\doibase
  10.1103/PhysRevB.92.174527} {\bibfield  {journal} {\bibinfo  {journal} {Phys.
  Rev. B}\ }\textbf {\bibinfo {volume} {92}},\ \bibinfo {pages} {174527}
  (\bibinfo {year} {2015})}\BibitemShut {NoStop}%
\bibitem [{\citenamefont {Fu}(2014)}]{Fu14}%
  \BibitemOpen
  \bibfield  {author} {\bibinfo {author} {\bibfnamefont {L.}~\bibnamefont
  {Fu}},\ }\href {\doibase 10.1103/PhysRevB.90.100509} {\bibfield  {journal}
  {\bibinfo  {journal} {Phys. Rev. B}\ }\textbf {\bibinfo {volume} {90}},\
  \bibinfo {pages} {100509} (\bibinfo {year} {2014})}\BibitemShut {NoStop}%
\bibitem [{\citenamefont {Matano}\ \emph {et~al.}(2016)\citenamefont {Matano},
  \citenamefont {Kriener}, \citenamefont {Segawa}, \citenamefont {Ando},\ and\
  \citenamefont {Zheng}}]{Matano16}%
  \BibitemOpen
  \bibfield  {author} {\bibinfo {author} {\bibfnamefont {K.}~\bibnamefont
  {Matano}}, \bibinfo {author} {\bibfnamefont {M.}~\bibnamefont {Kriener}},
  \bibinfo {author} {\bibfnamefont {K.}~\bibnamefont {Segawa}}, \bibinfo
  {author} {\bibfnamefont {Y.}~\bibnamefont {Ando}}, \ and\ \bibinfo {author}
  {\bibfnamefont {G.-q.}\ \bibnamefont {Zheng}},\ }\href {\doibase
  http://dx.doi.org/10.1038/nphys3781} {\bibfield  {journal} {\bibinfo
  {journal} {Nature Physics}\ }\textbf {\bibinfo {volume} {12}},\ \bibinfo
  {pages} {852} (\bibinfo {year} {2016})}\BibitemShut {NoStop}%
\bibitem [{\citenamefont {Yonezawa}\ \emph {et~al.}(2016)\citenamefont
  {Yonezawa}, \citenamefont {Tajiri}, \citenamefont {Nakata}, \citenamefont
  {Nagai}, \citenamefont {Wang}, \citenamefont {Segawa}, \citenamefont {Ando},\
  and\ \citenamefont {Maeno}}]{Yonezawa16}%
  \BibitemOpen
  \bibfield  {author} {\bibinfo {author} {\bibfnamefont {S.}~\bibnamefont
  {Yonezawa}}, \bibinfo {author} {\bibfnamefont {K.}~\bibnamefont {Tajiri}},
  \bibinfo {author} {\bibfnamefont {S.}~\bibnamefont {Nakata}}, \bibinfo
  {author} {\bibfnamefont {Y.}~\bibnamefont {Nagai}}, \bibinfo {author}
  {\bibfnamefont {Z.}~\bibnamefont {Wang}}, \bibinfo {author} {\bibfnamefont
  {K.}~\bibnamefont {Segawa}}, \bibinfo {author} {\bibfnamefont
  {Y.}~\bibnamefont {Ando}}, \ and\ \bibinfo {author} {\bibfnamefont
  {Y.}~\bibnamefont {Maeno}},\ }\href {\doibase
  http://dx.doi.org/10.1038/nphys3907} {\bibfield  {journal} {\bibinfo
  {journal} {Nature Physics}\ }\textbf {\bibinfo {volume} {13}},\ \bibinfo
  {pages} {123} (\bibinfo {year} {2016})}\BibitemShut {NoStop}%
\bibitem [{\citenamefont {Aggarwal}\ \emph {et~al.}(2015)\citenamefont
  {Aggarwal}, \citenamefont {Gaurav}, \citenamefont {Thakur}, \citenamefont
  {Haque}, \citenamefont {Ganguli},\ and\ \citenamefont {Sheet}}]{Aggarwal15}%
  \BibitemOpen
  \bibfield  {author} {\bibinfo {author} {\bibfnamefont {L.}~\bibnamefont
  {Aggarwal}}, \bibinfo {author} {\bibfnamefont {A.}~\bibnamefont {Gaurav}},
  \bibinfo {author} {\bibfnamefont {G.~S.}\ \bibnamefont {Thakur}}, \bibinfo
  {author} {\bibfnamefont {Z.}~\bibnamefont {Haque}}, \bibinfo {author}
  {\bibfnamefont {A.~K.}\ \bibnamefont {Ganguli}}, \ and\ \bibinfo {author}
  {\bibfnamefont {G.}~\bibnamefont {Sheet}},\ }\href {\doibase
  http://dx.doi.org/10.1038/nmat4455} {\bibfield  {journal} {\bibinfo
  {journal} {Nature Materials}\ }\textbf {\bibinfo {volume} {16}},\ \bibinfo
  {pages} {32–37} (\bibinfo {year} {2015})}\BibitemShut {NoStop}%
\bibitem [{\citenamefont {Wang}\ \emph {et~al.}(2015)\citenamefont {Wang},
  \citenamefont {Wang}, \citenamefont {Liu}, \citenamefont {Lu}, \citenamefont
  {Yang}, \citenamefont {Jia}, \citenamefont {Liu}, \citenamefont {Xie},
  \citenamefont {Wei},\ and\ \citenamefont {Wang}}]{Wang15}%
  \BibitemOpen
  \bibfield  {author} {\bibinfo {author} {\bibfnamefont {H.}~\bibnamefont
  {Wang}}, \bibinfo {author} {\bibfnamefont {H.}~\bibnamefont {Wang}}, \bibinfo
  {author} {\bibfnamefont {H.}~\bibnamefont {Liu}}, \bibinfo {author}
  {\bibfnamefont {H.}~\bibnamefont {Lu}}, \bibinfo {author} {\bibfnamefont
  {W.}~\bibnamefont {Yang}}, \bibinfo {author} {\bibfnamefont {S.}~\bibnamefont
  {Jia}}, \bibinfo {author} {\bibfnamefont {X.-J.}\ \bibnamefont {Liu}},
  \bibinfo {author} {\bibfnamefont {X.~C.}\ \bibnamefont {Xie}}, \bibinfo
  {author} {\bibfnamefont {J.}~\bibnamefont {Wei}}, \ and\ \bibinfo {author}
  {\bibfnamefont {J.}~\bibnamefont {Wang}},\ }\href {\doibase
  http://dx.doi.org/10.1038/nmat4456} {\bibfield  {journal} {\bibinfo
  {journal} {Nature Materials}\ }\textbf {\bibinfo {volume} {15}},\ \bibinfo
  {pages} {38–42} (\bibinfo {year} {2015})}\BibitemShut {NoStop}%
\bibitem [{\citenamefont {Kobayashi}\ and\ \citenamefont
  {Sato}(2015)}]{Kobayashi15PRL}%
  \BibitemOpen
  \bibfield  {author} {\bibinfo {author} {\bibfnamefont {S.}~\bibnamefont
  {Kobayashi}}\ and\ \bibinfo {author} {\bibfnamefont {M.}~\bibnamefont
  {Sato}},\ }\href {\doibase 10.1103/PhysRevLett.115.187001} {\bibfield
  {journal} {\bibinfo  {journal} {Phys. Rev. Lett.}\ }\textbf {\bibinfo
  {volume} {115}},\ \bibinfo {pages} {187001} (\bibinfo {year}
  {2015})}\BibitemShut {NoStop}%
\bibitem [{\citenamefont {Hashimoto}\ \emph {et~al.}(2016)\citenamefont
  {Hashimoto}, \citenamefont {Kobayashi}, \citenamefont {Tanaka},\ and\
  \citenamefont {Sato}}]{Hashimoto16}%
  \BibitemOpen
  \bibfield  {author} {\bibinfo {author} {\bibfnamefont {T.}~\bibnamefont
  {Hashimoto}}, \bibinfo {author} {\bibfnamefont {S.}~\bibnamefont
  {Kobayashi}}, \bibinfo {author} {\bibfnamefont {Y.}~\bibnamefont {Tanaka}}, \
  and\ \bibinfo {author} {\bibfnamefont {M.}~\bibnamefont {Sato}},\ }\href
  {\doibase 10.1103/PhysRevB.94.014510} {\bibfield  {journal} {\bibinfo
  {journal} {Phys. Rev. B}\ }\textbf {\bibinfo {volume} {94}},\ \bibinfo
  {pages} {014510} (\bibinfo {year} {2016})}\BibitemShut {NoStop}%
\bibitem [{\citenamefont {Oudah}\ \emph {et~al.}(2016)\citenamefont {Oudah},
  \citenamefont {Ikeda}, \citenamefont {Hausmann}, \citenamefont {Yonezawa},
  \citenamefont {Fukumoto}, \citenamefont {Kobayashi}, \citenamefont {Sato},\
  and\ \citenamefont {Maeno}}]{Oudah16}%
  \BibitemOpen
  \bibfield  {author} {\bibinfo {author} {\bibfnamefont {M.}~\bibnamefont
  {Oudah}}, \bibinfo {author} {\bibfnamefont {A.}~\bibnamefont {Ikeda}},
  \bibinfo {author} {\bibfnamefont {J.~N.}\ \bibnamefont {Hausmann}}, \bibinfo
  {author} {\bibfnamefont {S.}~\bibnamefont {Yonezawa}}, \bibinfo {author}
  {\bibfnamefont {T.}~\bibnamefont {Fukumoto}}, \bibinfo {author}
  {\bibfnamefont {S.}~\bibnamefont {Kobayashi}}, \bibinfo {author}
  {\bibfnamefont {M.}~\bibnamefont {Sato}}, \ and\ \bibinfo {author}
  {\bibfnamefont {Y.}~\bibnamefont {Maeno}},\ }\href {\doibase
  http://dx.doi.org/10.1038/ncomms13617} {\bibfield  {journal} {\bibinfo
  {journal} {Nature Communications}\ }\textbf {\bibinfo {volume} {7}},\
  \bibinfo {pages} {13617} (\bibinfo {year} {2016})}\BibitemShut {NoStop}%
\bibitem [{\citenamefont {Kawakami}\ \emph {et~al.}(2018)\citenamefont
  {Kawakami}, \citenamefont {Okamura}, \citenamefont {Kobayashi},\ and\
  \citenamefont {Sato}}]{Kawakami18}%
  \BibitemOpen
  \bibfield  {author} {\bibinfo {author} {\bibfnamefont {T.}~\bibnamefont
  {Kawakami}}, \bibinfo {author} {\bibfnamefont {T.}~\bibnamefont {Okamura}},
  \bibinfo {author} {\bibfnamefont {S.}~\bibnamefont {Kobayashi}}, \ and\
  \bibinfo {author} {\bibfnamefont {M.}~\bibnamefont {Sato}},\ }\href {\doibase
  10.1103/PhysRevX.8.041026} {\bibfield  {journal} {\bibinfo  {journal} {Phys.
  Rev. X}\ }\textbf {\bibinfo {volume} {8}},\ \bibinfo {pages} {041026}
  (\bibinfo {year} {2018})}\BibitemShut {NoStop}%
\bibitem [{\citenamefont {Majorana}(1937)}]{Majorana37}%
  \BibitemOpen
  \bibfield  {author} {\bibinfo {author} {\bibfnamefont {E.}~\bibnamefont
  {Majorana}},\ }\href {http://dx.doi.org/10.1007/BF02961314} {\bibfield
  {journal} {\bibinfo  {journal} {Nuovo Cimento}\ }\textbf {\bibinfo {volume}
  {14}},\ \bibinfo {pages} {171} (\bibinfo {year} {1937})}\BibitemShut
  {NoStop}%
\bibitem [{\citenamefont {Avignone}\ \emph {et~al.}(2008)\citenamefont
  {Avignone}, \citenamefont {Elliott},\ and\ \citenamefont
  {Engel}}]{Avignone08}%
  \BibitemOpen
  \bibfield  {author} {\bibinfo {author} {\bibfnamefont {F.~T.}\ \bibnamefont
  {Avignone}}, \bibinfo {author} {\bibfnamefont {S.~R.}\ \bibnamefont
  {Elliott}}, \ and\ \bibinfo {author} {\bibfnamefont {J.}~\bibnamefont
  {Engel}},\ }\href {\doibase 10.1103/RevModPhys.80.481} {\bibfield  {journal}
  {\bibinfo  {journal} {Rev. Mod. Phys.}\ }\textbf {\bibinfo {volume} {80}},\
  \bibinfo {pages} {481} (\bibinfo {year} {2008})}\BibitemShut {NoStop}%
\bibitem [{\citenamefont {Kayser}\ and\ \citenamefont
  {Goldhaber}(1983)}]{Kayser83}%
  \BibitemOpen
  \bibfield  {author} {\bibinfo {author} {\bibfnamefont {B.}~\bibnamefont
  {Kayser}}\ and\ \bibinfo {author} {\bibfnamefont {A.~S.}\ \bibnamefont
  {Goldhaber}},\ }\href {\doibase 10.1103/PhysRevD.28.2341} {\bibfield
  {journal} {\bibinfo  {journal} {Phys. Rev. D}\ }\textbf {\bibinfo {volume}
  {28}},\ \bibinfo {pages} {2341} (\bibinfo {year} {1983})}\BibitemShut
  {NoStop}%
\bibitem [{\citenamefont {Radescu}(1985)}]{Radescu85}%
  \BibitemOpen
  \bibfield  {author} {\bibinfo {author} {\bibfnamefont {E.~E.}\ \bibnamefont
  {Radescu}},\ }\href {\doibase 10.1103/PhysRevD.32.1266} {\bibfield  {journal}
  {\bibinfo  {journal} {Phys. Rev. D}\ }\textbf {\bibinfo {volume} {32}},\
  \bibinfo {pages} {1266} (\bibinfo {year} {1985})}\BibitemShut {NoStop}%
\bibitem [{\citenamefont {Boudjema}\ \emph {et~al.}(1989)\citenamefont
  {Boudjema}, \citenamefont {Hamzaoui}, \citenamefont {Rahal},\ and\
  \citenamefont {Ren}}]{Boudjema89}%
  \BibitemOpen
  \bibfield  {author} {\bibinfo {author} {\bibfnamefont {F.}~\bibnamefont
  {Boudjema}}, \bibinfo {author} {\bibfnamefont {C.}~\bibnamefont {Hamzaoui}},
  \bibinfo {author} {\bibfnamefont {V.}~\bibnamefont {Rahal}}, \ and\ \bibinfo
  {author} {\bibfnamefont {H.~C.}\ \bibnamefont {Ren}},\ }\href {\doibase
  10.1103/PhysRevLett.62.852} {\bibfield  {journal} {\bibinfo  {journal} {Phys.
  Rev. Lett.}\ }\textbf {\bibinfo {volume} {62}},\ \bibinfo {pages} {852}
  (\bibinfo {year} {1989})}\BibitemShut {NoStop}%
\bibitem [{\citenamefont {Goll}\ \emph {et~al.}(2008)\citenamefont {Goll},
  \citenamefont {Marz}, \citenamefont {Hamann}, \citenamefont {Tomanic},
  \citenamefont {Grube}, \citenamefont {Yoshino},\ and\ \citenamefont
  {Takabatake}}]{Goll08}%
  \BibitemOpen
  \bibfield  {author} {\bibinfo {author} {\bibfnamefont {G.}~\bibnamefont
  {Goll}}, \bibinfo {author} {\bibfnamefont {M.}~\bibnamefont {Marz}}, \bibinfo
  {author} {\bibfnamefont {A.}~\bibnamefont {Hamann}}, \bibinfo {author}
  {\bibfnamefont {T.}~\bibnamefont {Tomanic}}, \bibinfo {author} {\bibfnamefont
  {K.}~\bibnamefont {Grube}}, \bibinfo {author} {\bibfnamefont
  {T.}~\bibnamefont {Yoshino}}, \ and\ \bibinfo {author} {\bibfnamefont
  {T.}~\bibnamefont {Takabatake}},\ }\href {\doibase
  https://doi.org/10.1016/j.physb.2007.10.089} {\bibfield  {journal} {\bibinfo
  {journal} {Physica B: Condensed Matter}\ }\textbf {\bibinfo {volume} {403}},\
  \bibinfo {pages} {1065 } (\bibinfo {year} {2008})}\BibitemShut {NoStop}%
\bibitem [{\citenamefont {Butch}\ \emph {et~al.}(2011)\citenamefont {Butch},
  \citenamefont {Syers}, \citenamefont {Kirshenbaum}, \citenamefont {Hope},\
  and\ \citenamefont {Paglione}}]{Butch11}%
  \BibitemOpen
  \bibfield  {author} {\bibinfo {author} {\bibfnamefont {N.~P.}\ \bibnamefont
  {Butch}}, \bibinfo {author} {\bibfnamefont {P.}~\bibnamefont {Syers}},
  \bibinfo {author} {\bibfnamefont {K.}~\bibnamefont {Kirshenbaum}}, \bibinfo
  {author} {\bibfnamefont {A.~P.}\ \bibnamefont {Hope}}, \ and\ \bibinfo
  {author} {\bibfnamefont {J.}~\bibnamefont {Paglione}},\ }\href {\doibase
  10.1103/PhysRevB.84.220504} {\bibfield  {journal} {\bibinfo  {journal} {Phys.
  Rev. B}\ }\textbf {\bibinfo {volume} {84}},\ \bibinfo {pages} {220504}
  (\bibinfo {year} {2011})}\BibitemShut {NoStop}%
\bibitem [{\citenamefont {Tafti}\ \emph {et~al.}(2013)\citenamefont {Tafti},
  \citenamefont {Fujii}, \citenamefont {Juneau-Fecteau}, \citenamefont
  {Ren\'e~de Cotret}, \citenamefont {Doiron-Leyraud}, \citenamefont
  {Asamitsu},\ and\ \citenamefont {Taillefer}}]{Tafti13}%
  \BibitemOpen
  \bibfield  {author} {\bibinfo {author} {\bibfnamefont {F.~F.}\ \bibnamefont
  {Tafti}}, \bibinfo {author} {\bibfnamefont {T.}~\bibnamefont {Fujii}},
  \bibinfo {author} {\bibfnamefont {A.}~\bibnamefont {Juneau-Fecteau}},
  \bibinfo {author} {\bibfnamefont {S.}~\bibnamefont {Ren\'e~de Cotret}},
  \bibinfo {author} {\bibfnamefont {N.}~\bibnamefont {Doiron-Leyraud}},
  \bibinfo {author} {\bibfnamefont {A.}~\bibnamefont {Asamitsu}}, \ and\
  \bibinfo {author} {\bibfnamefont {L.}~\bibnamefont {Taillefer}},\ }\href
  {\doibase 10.1103/PhysRevB.87.184504} {\bibfield  {journal} {\bibinfo
  {journal} {Phys. Rev. B}\ }\textbf {\bibinfo {volume} {87}},\ \bibinfo
  {pages} {184504} (\bibinfo {year} {2013})}\BibitemShut {NoStop}%
\bibitem [{\citenamefont {Xu}\ \emph {et~al.}(2014)\citenamefont {Xu},
  \citenamefont {Wang}, \citenamefont {Zhang}, \citenamefont {Du},
  \citenamefont {Liu}, \citenamefont {Wang}, \citenamefont {Wu}, \citenamefont
  {Liu},\ and\ \citenamefont {Zhang}}]{GXu16}%
  \BibitemOpen
  \bibfield  {author} {\bibinfo {author} {\bibfnamefont {G.}~\bibnamefont
  {Xu}}, \bibinfo {author} {\bibfnamefont {W.}~\bibnamefont {Wang}}, \bibinfo
  {author} {\bibfnamefont {X.}~\bibnamefont {Zhang}}, \bibinfo {author}
  {\bibfnamefont {Y.}~\bibnamefont {Du}}, \bibinfo {author} {\bibfnamefont
  {E.}~\bibnamefont {Liu}}, \bibinfo {author} {\bibfnamefont {S.}~\bibnamefont
  {Wang}}, \bibinfo {author} {\bibfnamefont {G.}~\bibnamefont {Wu}}, \bibinfo
  {author} {\bibfnamefont {Z.}~\bibnamefont {Liu}}, \ and\ \bibinfo {author}
  {\bibfnamefont {X.~X.}\ \bibnamefont {Zhang}},\ }\href {\doibase
  http://dx.doi.org/10.1038/srep05709} {\bibfield  {journal} {\bibinfo
  {journal} {Scientific Reports}\ }\textbf {\bibinfo {volume} {4}},\ \bibinfo
  {pages} {5709} (\bibinfo {year} {2014})}\BibitemShut {NoStop}%
\bibitem [{\citenamefont {Bay}\ \emph {et~al.}(2012)\citenamefont {Bay},
  \citenamefont {Naka}, \citenamefont {Huang},\ and\ \citenamefont
  {de~Visser}}]{Bay12}%
  \BibitemOpen
  \bibfield  {author} {\bibinfo {author} {\bibfnamefont {T.~V.}\ \bibnamefont
  {Bay}}, \bibinfo {author} {\bibfnamefont {T.}~\bibnamefont {Naka}}, \bibinfo
  {author} {\bibfnamefont {Y.~K.}\ \bibnamefont {Huang}}, \ and\ \bibinfo
  {author} {\bibfnamefont {A.}~\bibnamefont {de~Visser}},\ }\href {\doibase
  10.1103/PhysRevB.86.064515} {\bibfield  {journal} {\bibinfo  {journal} {Phys.
  Rev. B}\ }\textbf {\bibinfo {volume} {86}},\ \bibinfo {pages} {064515}
  (\bibinfo {year} {2012})}\BibitemShut {NoStop}%
\bibitem [{\citenamefont {Kim}\ \emph {et~al.}(2018)\citenamefont {Kim},
  \citenamefont {Wang}, \citenamefont {Nakajima}, \citenamefont {Hu},
  \citenamefont {Ziemak}, \citenamefont {Syers}, \citenamefont {Wang},
  \citenamefont {Hodovanets}, \citenamefont {Denlinger}, \citenamefont
  {Brydon}, \citenamefont {Agterberg}, \citenamefont {Tanatar}, \citenamefont
  {Prozorov},\ and\ \citenamefont {Paglione}}]{Kim18}%
  \BibitemOpen
  \bibfield  {author} {\bibinfo {author} {\bibfnamefont {H.}~\bibnamefont
  {Kim}}, \bibinfo {author} {\bibfnamefont {K.}~\bibnamefont {Wang}}, \bibinfo
  {author} {\bibfnamefont {Y.}~\bibnamefont {Nakajima}}, \bibinfo {author}
  {\bibfnamefont {R.}~\bibnamefont {Hu}}, \bibinfo {author} {\bibfnamefont
  {S.}~\bibnamefont {Ziemak}}, \bibinfo {author} {\bibfnamefont
  {P.}~\bibnamefont {Syers}}, \bibinfo {author} {\bibfnamefont
  {L.}~\bibnamefont {Wang}}, \bibinfo {author} {\bibfnamefont {H.}~\bibnamefont
  {Hodovanets}}, \bibinfo {author} {\bibfnamefont {J.~D.}\ \bibnamefont
  {Denlinger}}, \bibinfo {author} {\bibfnamefont {P.~M.~R.}\ \bibnamefont
  {Brydon}}, \bibinfo {author} {\bibfnamefont {D.~F.}\ \bibnamefont
  {Agterberg}}, \bibinfo {author} {\bibfnamefont {M.~A.}\ \bibnamefont
  {Tanatar}}, \bibinfo {author} {\bibfnamefont {R.}~\bibnamefont {Prozorov}}, \
  and\ \bibinfo {author} {\bibfnamefont {J.}~\bibnamefont {Paglione}},\ }\href
  {\doibase 10.1126/sciadv.aao4513} {\bibfield  {journal} {\bibinfo  {journal}
  {Science Advances}\ }\textbf {\bibinfo {volume} {4}},\ \bibinfo {pages}
  {eaao4513} (\bibinfo {year} {2018})}\BibitemShut {NoStop}%
\bibitem [{\citenamefont {Boettcher}\ and\ \citenamefont
  {Herbut}(2016)}]{Boettcher16}%
  \BibitemOpen
  \bibfield  {author} {\bibinfo {author} {\bibfnamefont {I.}~\bibnamefont
  {Boettcher}}\ and\ \bibinfo {author} {\bibfnamefont {I.~F.}\ \bibnamefont
  {Herbut}},\ }\href {\doibase 10.1103/PhysRevB.93.205138} {\bibfield
  {journal} {\bibinfo  {journal} {Phys. Rev. B}\ }\textbf {\bibinfo {volume}
  {93}},\ \bibinfo {pages} {205138} (\bibinfo {year} {2016})}\BibitemShut
  {NoStop}%
\bibitem [{\citenamefont {Brydon}\ \emph {et~al.}(2016)\citenamefont {Brydon},
  \citenamefont {Wang}, \citenamefont {Weinert},\ and\ \citenamefont
  {Agterberg}}]{Brydon16}%
  \BibitemOpen
  \bibfield  {author} {\bibinfo {author} {\bibfnamefont {P.~M.~R.}\
  \bibnamefont {Brydon}}, \bibinfo {author} {\bibfnamefont {L.}~\bibnamefont
  {Wang}}, \bibinfo {author} {\bibfnamefont {M.}~\bibnamefont {Weinert}}, \
  and\ \bibinfo {author} {\bibfnamefont {D.~F.}\ \bibnamefont {Agterberg}},\
  }\href {\doibase 10.1103/PhysRevLett.116.177001} {\bibfield  {journal}
  {\bibinfo  {journal} {Phys. Rev. Lett.}\ }\textbf {\bibinfo {volume} {116}},\
  \bibinfo {pages} {177001} (\bibinfo {year} {2016})}\BibitemShut {NoStop}%
\bibitem [{\citenamefont {Savary}\ \emph {et~al.}(2017)\citenamefont {Savary},
  \citenamefont {Ruhman}, \citenamefont {Venderbos}, \citenamefont {Fu},\ and\
  \citenamefont {Lee}}]{Savary17}%
  \BibitemOpen
  \bibfield  {author} {\bibinfo {author} {\bibfnamefont {L.}~\bibnamefont
  {Savary}}, \bibinfo {author} {\bibfnamefont {J.}~\bibnamefont {Ruhman}},
  \bibinfo {author} {\bibfnamefont {J.~W.~F.}\ \bibnamefont {Venderbos}},
  \bibinfo {author} {\bibfnamefont {L.}~\bibnamefont {Fu}}, \ and\ \bibinfo
  {author} {\bibfnamefont {P.~A.}\ \bibnamefont {Lee}},\ }\href {\doibase
  10.1103/PhysRevB.96.214514} {\bibfield  {journal} {\bibinfo  {journal} {Phys.
  Rev. B}\ }\textbf {\bibinfo {volume} {96}},\ \bibinfo {pages} {214514}
  (\bibinfo {year} {2017})}\BibitemShut {NoStop}%
\bibitem [{\citenamefont {{Roy}}\ \emph {et~al.}(2017)\citenamefont {{Roy}},
  \citenamefont {{Ghorashi}}, \citenamefont {{Foster}},\ and\ \citenamefont
  {{Nevidomskyy}}}]{Roy17}%
  \BibitemOpen
  \bibfield  {author} {\bibinfo {author} {\bibfnamefont {B.}~\bibnamefont
  {{Roy}}}, \bibinfo {author} {\bibfnamefont {S.~A.~A.}\ \bibnamefont
  {{Ghorashi}}}, \bibinfo {author} {\bibfnamefont {M.~S.}\ \bibnamefont
  {{Foster}}}, \ and\ \bibinfo {author} {\bibfnamefont {A.~H.}\ \bibnamefont
  {{Nevidomskyy}}},\ }\href@noop {} {\bibfield  {journal} {\bibinfo  {journal}
  {ArXiv e-prints}\ } (\bibinfo {year} {2017})},\ \Eprint
  {http://arxiv.org/abs/1708.07825} {arXiv:1708.07825 [cond-mat.mes-hall]}
  \BibitemShut {NoStop}%
\bibitem [{\citenamefont {Timm}\ \emph {et~al.}(2017)\citenamefont {Timm},
  \citenamefont {Schnyder}, \citenamefont {Agterberg},\ and\ \citenamefont
  {Brydon}}]{Timm17}%
  \BibitemOpen
  \bibfield  {author} {\bibinfo {author} {\bibfnamefont {C.}~\bibnamefont
  {Timm}}, \bibinfo {author} {\bibfnamefont {A.~P.}\ \bibnamefont {Schnyder}},
  \bibinfo {author} {\bibfnamefont {D.~F.}\ \bibnamefont {Agterberg}}, \ and\
  \bibinfo {author} {\bibfnamefont {P.~M.~R.}\ \bibnamefont {Brydon}},\ }\href
  {\doibase 10.1103/PhysRevB.96.094526} {\bibfield  {journal} {\bibinfo
  {journal} {Phys. Rev. B}\ }\textbf {\bibinfo {volume} {96}},\ \bibinfo
  {pages} {094526} (\bibinfo {year} {2017})}\BibitemShut {NoStop}%
\bibitem [{\citenamefont {Venderbos}\ \emph {et~al.}(2018)\citenamefont
  {Venderbos}, \citenamefont {Savary}, \citenamefont {Ruhman}, \citenamefont
  {Lee},\ and\ \citenamefont {Fu}}]{Venderbos18}%
  \BibitemOpen
  \bibfield  {author} {\bibinfo {author} {\bibfnamefont {J.~W.~F.}\
  \bibnamefont {Venderbos}}, \bibinfo {author} {\bibfnamefont {L.}~\bibnamefont
  {Savary}}, \bibinfo {author} {\bibfnamefont {J.}~\bibnamefont {Ruhman}},
  \bibinfo {author} {\bibfnamefont {P.~A.}\ \bibnamefont {Lee}}, \ and\
  \bibinfo {author} {\bibfnamefont {L.}~\bibnamefont {Fu}},\ }\href {\doibase
  10.1103/PhysRevX.8.011029} {\bibfield  {journal} {\bibinfo  {journal} {Phys.
  Rev. X}\ }\textbf {\bibinfo {volume} {8}},\ \bibinfo {pages} {011029}
  (\bibinfo {year} {2018})}\BibitemShut {NoStop}%
\bibitem [{\citenamefont {Boettcher}\ and\ \citenamefont
  {Herbut}(2018)}]{Boettcher18}%
  \BibitemOpen
  \bibfield  {author} {\bibinfo {author} {\bibfnamefont {I.}~\bibnamefont
  {Boettcher}}\ and\ \bibinfo {author} {\bibfnamefont {I.~F.}\ \bibnamefont
  {Herbut}},\ }\href {\doibase 10.1103/PhysRevLett.120.057002} {\bibfield
  {journal} {\bibinfo  {journal} {Phys. Rev. Lett.}\ }\textbf {\bibinfo
  {volume} {120}},\ \bibinfo {pages} {057002} (\bibinfo {year}
  {2018})}\BibitemShut {NoStop}%
\bibitem [{\citenamefont {{Yu}}\ and\ \citenamefont {{Liu}}(2018)}]{Yu18}%
  \BibitemOpen
  \bibfield  {author} {\bibinfo {author} {\bibfnamefont {J.}~\bibnamefont
  {{Yu}}}\ and\ \bibinfo {author} {\bibfnamefont {C.-X.}\ \bibnamefont
  {{Liu}}},\ }\href@noop {} {\bibfield  {journal} {\bibinfo  {journal} {ArXiv
  e-prints}\ } (\bibinfo {year} {2018})},\ \Eprint
  {http://arxiv.org/abs/1801.00083} {arXiv:1801.00083 [cond-mat.supr-con]}
  \BibitemShut {NoStop}%
\bibitem [{\citenamefont {Kuzmenko}\ \emph {et~al.}(2018)\citenamefont
  {Kuzmenko}, \citenamefont {Kuzmenko}, \citenamefont {Avishai},\ and\
  \citenamefont {Sato}}]{Kuzmenko18}%
  \BibitemOpen
  \bibfield  {author} {\bibinfo {author} {\bibfnamefont {I.}~\bibnamefont
  {Kuzmenko}}, \bibinfo {author} {\bibfnamefont {T.}~\bibnamefont {Kuzmenko}},
  \bibinfo {author} {\bibfnamefont {Y.}~\bibnamefont {Avishai}}, \ and\
  \bibinfo {author} {\bibfnamefont {M.}~\bibnamefont {Sato}},\ }\href {\doibase
  10.1103/PhysRevB.98.165139} {\bibfield  {journal} {\bibinfo  {journal} {Phys.
  Rev. B}\ }\textbf {\bibinfo {volume} {98}},\ \bibinfo {pages} {165139}
  (\bibinfo {year} {2018})}\BibitemShut {NoStop}%
\bibitem [{\citenamefont {Grinevich}\ and\ \citenamefont
  {Volovik}(1988)}]{Grinevich88}%
  \BibitemOpen
  \bibfield  {author} {\bibinfo {author} {\bibfnamefont {P.~G.}\ \bibnamefont
  {Grinevich}}\ and\ \bibinfo {author} {\bibfnamefont {G.~E.}\ \bibnamefont
  {Volovik}},\ }\href {\doibase 10.1007/BF00682148} {\bibfield  {journal}
  {\bibinfo  {journal} {Journal of Low Temperature Physics}\ }\textbf {\bibinfo
  {volume} {72}},\ \bibinfo {pages} {371} (\bibinfo {year} {1988})}\BibitemShut
  {NoStop}%
\bibitem [{\citenamefont {Sato}(2010)}]{Sato10}%
  \BibitemOpen
  \bibfield  {author} {\bibinfo {author} {\bibfnamefont {M.}~\bibnamefont
  {Sato}},\ }\href {\doibase 10.1103/PhysRevB.81.220504} {\bibfield  {journal}
  {\bibinfo  {journal} {Phys. Rev. B}\ }\textbf {\bibinfo {volume} {81}},\
  \bibinfo {pages} {220504} (\bibinfo {year} {2010})}\BibitemShut {NoStop}%
\bibitem [{Note1()}]{note1}%
  \BibitemOpen
  \bibinfo {note} {These relations lead to $\protect \{\protect \mathcal {C},
  \rho ^{(12)}\protect \}=\protect \{\protect \mathcal {T}, \rho
  ^{(12)}\protect \}=0$, which are consistent with $\protect \{{\protect \cal
  C}, {\protect \cal O}_{\Gamma }\protect \}=\protect \{{\protect \cal T},
  {\protect \cal O}_{\Gamma }\protect \}=0$.}\BibitemShut {Stop}%
\bibitem [{\citenamefont {Mizushima}\ \emph {et~al.}(2012)\citenamefont
  {Mizushima}, \citenamefont {Sato},\ and\ \citenamefont
  {Machida}}]{Mizushima12}%
  \BibitemOpen
  \bibfield  {author} {\bibinfo {author} {\bibfnamefont {T.}~\bibnamefont
  {Mizushima}}, \bibinfo {author} {\bibfnamefont {M.}~\bibnamefont {Sato}}, \
  and\ \bibinfo {author} {\bibfnamefont {K.}~\bibnamefont {Machida}},\ }\href
  {\doibase 10.1103/PhysRevLett.109.165301} {\bibfield  {journal} {\bibinfo
  {journal} {Phys. Rev. Lett.}\ }\textbf {\bibinfo {volume} {109}},\ \bibinfo
  {pages} {165301} (\bibinfo {year} {2012})}\BibitemShut {NoStop}%
\bibitem [{\citenamefont {Shiozaki}\ and\ \citenamefont
  {Sato}(2014)}]{Shiozaki14}%
  \BibitemOpen
  \bibfield  {author} {\bibinfo {author} {\bibfnamefont {K.}~\bibnamefont
  {Shiozaki}}\ and\ \bibinfo {author} {\bibfnamefont {M.}~\bibnamefont
  {Sato}},\ }\href {\doibase 10.1103/PhysRevB.90.165114} {\bibfield  {journal}
  {\bibinfo  {journal} {Phys. Rev. B}\ }\textbf {\bibinfo {volume} {90}},\
  \bibinfo {pages} {165114} (\bibinfo {year} {2014})}\BibitemShut {NoStop}%
\bibitem [{\citenamefont {Dumitrescu}\ \emph {et~al.}(2014)\citenamefont
  {Dumitrescu}, \citenamefont {Sau},\ and\ \citenamefont
  {Tewari}}]{Dumitrescu2014}%
  \BibitemOpen
  \bibfield  {author} {\bibinfo {author} {\bibfnamefont {E.}~\bibnamefont
  {Dumitrescu}}, \bibinfo {author} {\bibfnamefont {J.~D.}\ \bibnamefont {Sau}},
  \ and\ \bibinfo {author} {\bibfnamefont {S.}~\bibnamefont {Tewari}},\ }\href
  {\doibase 10.1103/PhysRevB.90.245438} {\bibfield  {journal} {\bibinfo
  {journal} {Phys. Rev. B}\ }\textbf {\bibinfo {volume} {90}},\ \bibinfo
  {pages} {245438} (\bibinfo {year} {2014})}\BibitemShut {NoStop}%
\bibitem [{\citenamefont {Xiong}\ \emph {et~al.}(2017)\citenamefont {Xiong},
  \citenamefont {Yamakage}, \citenamefont {Kobayashi}, \citenamefont {Sato},\
  and\ \citenamefont {Tanaka}}]{Xiong17}%
  \BibitemOpen
  \bibfield  {author} {\bibinfo {author} {\bibfnamefont {Y.}~\bibnamefont
  {Xiong}}, \bibinfo {author} {\bibfnamefont {A.}~\bibnamefont {Yamakage}},
  \bibinfo {author} {\bibfnamefont {S.}~\bibnamefont {Kobayashi}}, \bibinfo
  {author} {\bibfnamefont {M.}~\bibnamefont {Sato}}, \ and\ \bibinfo {author}
  {\bibfnamefont {Y.}~\bibnamefont {Tanaka}},\ }\href {\doibase
  10.3390/cryst7020058} {\bibfield  {journal} {\bibinfo  {journal} {Crystals}\
  }\textbf {\bibinfo {volume} {7}},\ \bibinfo {pages} {58} (\bibinfo {year}
  {2017})}\BibitemShut {NoStop}%
\bibitem [{sup()}]{suppl}%
  \BibitemOpen
  \href@noop {} {}\bibinfo {note} {See the Supplemental Material at [ URL will
  be inserted by publisher] for the detail calculation.}\BibitemShut {Stop}%
\bibitem [{\citenamefont {Sato}\ \emph {et~al.}(2011)\citenamefont {Sato},
  \citenamefont {Tanaka}, \citenamefont {Yada},\ and\ \citenamefont
  {Yokoyama}}]{Sato11}%
  \BibitemOpen
  \bibfield  {author} {\bibinfo {author} {\bibfnamefont {M.}~\bibnamefont
  {Sato}}, \bibinfo {author} {\bibfnamefont {Y.}~\bibnamefont {Tanaka}},
  \bibinfo {author} {\bibfnamefont {K.}~\bibnamefont {Yada}}, \ and\ \bibinfo
  {author} {\bibfnamefont {T.}~\bibnamefont {Yokoyama}},\ }\href {\doibase
  10.1103/PhysRevB.83.224511} {\bibfield  {journal} {\bibinfo  {journal} {Phys.
  Rev. B}\ }\textbf {\bibinfo {volume} {83}},\ \bibinfo {pages} {224511}
  (\bibinfo {year} {2011})}\BibitemShut {NoStop}%
\bibitem [{\citenamefont {Ho}\ and\ \citenamefont {Yip}(1999)}]{Ho99}%
  \BibitemOpen
  \bibfield  {author} {\bibinfo {author} {\bibfnamefont {T.-L.}\ \bibnamefont
  {Ho}}\ and\ \bibinfo {author} {\bibfnamefont {S.}~\bibnamefont {Yip}},\
  }\href {\doibase 10.1103/PhysRevLett.82.247} {\bibfield  {journal} {\bibinfo
  {journal} {Phys. Rev. Lett.}\ }\textbf {\bibinfo {volume} {82}},\ \bibinfo
  {pages} {247} (\bibinfo {year} {1999})}\BibitemShut {NoStop}%
\bibitem [{\citenamefont {Yang}\ \emph {et~al.}(2016)\citenamefont {Yang},
  \citenamefont {Li},\ and\ \citenamefont {Wu}}]{WYang16}%
  \BibitemOpen
  \bibfield  {author} {\bibinfo {author} {\bibfnamefont {W.}~\bibnamefont
  {Yang}}, \bibinfo {author} {\bibfnamefont {Y.}~\bibnamefont {Li}}, \ and\
  \bibinfo {author} {\bibfnamefont {C.}~\bibnamefont {Wu}},\ }\href {\doibase
  10.1103/PhysRevLett.117.075301} {\bibfield  {journal} {\bibinfo  {journal}
  {Phys. Rev. Lett.}\ }\textbf {\bibinfo {volume} {117}},\ \bibinfo {pages}
  {075301} (\bibinfo {year} {2016})}\BibitemShut {NoStop}%
\bibitem [{\citenamefont {Gor'kov}\ and\ \citenamefont
  {Rashba}(2001)}]{Gorkov01}%
  \BibitemOpen
  \bibfield  {author} {\bibinfo {author} {\bibfnamefont {L.~P.}\ \bibnamefont
  {Gor'kov}}\ and\ \bibinfo {author} {\bibfnamefont {E.~I.}\ \bibnamefont
  {Rashba}},\ }\href {\doibase 10.1103/PhysRevLett.87.037004} {\bibfield
  {journal} {\bibinfo  {journal} {Phys. Rev. Lett.}\ }\textbf {\bibinfo
  {volume} {87}},\ \bibinfo {pages} {037004} (\bibinfo {year}
  {2001})}\BibitemShut {NoStop}%
\bibitem [{\citenamefont {Frigeri}\ \emph {et~al.}(2004)\citenamefont
  {Frigeri}, \citenamefont {Agterberg}, \citenamefont {Koga},\ and\
  \citenamefont {Sigrist}}]{Frigeri04}%
  \BibitemOpen
  \bibfield  {author} {\bibinfo {author} {\bibfnamefont {P.~A.}\ \bibnamefont
  {Frigeri}}, \bibinfo {author} {\bibfnamefont {D.~F.}\ \bibnamefont
  {Agterberg}}, \bibinfo {author} {\bibfnamefont {A.}~\bibnamefont {Koga}}, \
  and\ \bibinfo {author} {\bibfnamefont {M.}~\bibnamefont {Sigrist}},\ }\href
  {\doibase 10.1103/PhysRevLett.92.097001} {\bibfield  {journal} {\bibinfo
  {journal} {Phys. Rev. Lett.}\ }\textbf {\bibinfo {volume} {92}},\ \bibinfo
  {pages} {097001} (\bibinfo {year} {2004})}\BibitemShut {NoStop}%
\bibitem [{\citenamefont {Fujimoto}(2007{\natexlab{a}})}]{Fujimoto07a}%
  \BibitemOpen
  \bibfield  {author} {\bibinfo {author} {\bibfnamefont {S.}~\bibnamefont
  {Fujimoto}},\ }\href {\doibase 10.1143/JPSJ.76.034712} {\bibfield  {journal}
  {\bibinfo  {journal} {Journal of the Physical Society of Japan}\ }\textbf
  {\bibinfo {volume} {76}},\ \bibinfo {pages} {034712} (\bibinfo {year}
  {2007}{\natexlab{a}})}\BibitemShut {NoStop}%
\bibitem [{\citenamefont {Fujimoto}(2007{\natexlab{b}})}]{Fujimoto07b}%
  \BibitemOpen
  \bibfield  {author} {\bibinfo {author} {\bibfnamefont {S.}~\bibnamefont
  {Fujimoto}},\ }\href {\doibase 10.1143/JPSJ.76.051008} {\bibfield  {journal}
  {\bibinfo  {journal} {Journal of the Physical Society of Japan}\ }\textbf
  {\bibinfo {volume} {76}},\ \bibinfo {pages} {051008} (\bibinfo {year}
  {2007}{\natexlab{b}})}\BibitemShut {NoStop}%
\bibitem [{\citenamefont {Bauer}\ \emph {et~al.}(2012)\citenamefont {Bauer},
  \citenamefont {Sigrist},\ and\ \citenamefont {editors}}]{Bauer12}%
  \BibitemOpen
  \bibfield  {author} {\bibinfo {author} {\bibfnamefont {E.}~\bibnamefont
  {Bauer}}, \bibinfo {author} {\bibfnamefont {M.}~\bibnamefont {Sigrist}}, \
  and\ \bibinfo {author} {\bibnamefont {editors}},\ }\href@noop {} {\emph
  {\bibinfo {title} {Non-Centrosymmetric Superconductors: Introduction and
  Overview}}}\ (\bibinfo  {publisher} {Springer},\ \bibinfo {address}
  {Heidelberg},\ \bibinfo {year} {2012})\BibitemShut {NoStop}%
\bibitem [{\citenamefont {Tanaka}\ \emph {et~al.}(2010)\citenamefont {Tanaka},
  \citenamefont {Mizuno}, \citenamefont {Yokoyama}, \citenamefont {Yada},\ and\
  \citenamefont {Sato}}]{Tanaka10}%
  \BibitemOpen
  \bibfield  {author} {\bibinfo {author} {\bibfnamefont {Y.}~\bibnamefont
  {Tanaka}}, \bibinfo {author} {\bibfnamefont {Y.}~\bibnamefont {Mizuno}},
  \bibinfo {author} {\bibfnamefont {T.}~\bibnamefont {Yokoyama}}, \bibinfo
  {author} {\bibfnamefont {K.}~\bibnamefont {Yada}}, \ and\ \bibinfo {author}
  {\bibfnamefont {M.}~\bibnamefont {Sato}},\ }\href {\doibase
  10.1103/PhysRevLett.105.097002} {\bibfield  {journal} {\bibinfo  {journal}
  {Phys. Rev. Lett.}\ }\textbf {\bibinfo {volume} {105}},\ \bibinfo {pages}
  {097002} (\bibinfo {year} {2010})}\BibitemShut {NoStop}%
\bibitem [{\citenamefont {Yada}\ \emph {et~al.}(2011)\citenamefont {Yada},
  \citenamefont {Sato}, \citenamefont {Tanaka},\ and\ \citenamefont
  {Yokoyama}}]{Yada11}%
  \BibitemOpen
  \bibfield  {author} {\bibinfo {author} {\bibfnamefont {K.}~\bibnamefont
  {Yada}}, \bibinfo {author} {\bibfnamefont {M.}~\bibnamefont {Sato}}, \bibinfo
  {author} {\bibfnamefont {Y.}~\bibnamefont {Tanaka}}, \ and\ \bibinfo {author}
  {\bibfnamefont {T.}~\bibnamefont {Yokoyama}},\ }\href {\doibase
  10.1103/PhysRevB.83.064505} {\bibfield  {journal} {\bibinfo  {journal} {Phys.
  Rev. B}\ }\textbf {\bibinfo {volume} {83}},\ \bibinfo {pages} {064505}
  (\bibinfo {year} {2011})}\BibitemShut {NoStop}%
\bibitem [{\citenamefont {Brydon}\ \emph {et~al.}(2011)\citenamefont {Brydon},
  \citenamefont {Schnyder},\ and\ \citenamefont {Timm}}]{Brydon11}%
  \BibitemOpen
  \bibfield  {author} {\bibinfo {author} {\bibfnamefont {P.~M.~R.}\
  \bibnamefont {Brydon}}, \bibinfo {author} {\bibfnamefont {A.~P.}\
  \bibnamefont {Schnyder}}, \ and\ \bibinfo {author} {\bibfnamefont
  {C.}~\bibnamefont {Timm}},\ }\href {\doibase 10.1103/PhysRevB.84.020501}
  {\bibfield  {journal} {\bibinfo  {journal} {Phys. Rev. B}\ }\textbf {\bibinfo
  {volume} {84}},\ \bibinfo {pages} {020501} (\bibinfo {year}
  {2011})}\BibitemShut {NoStop}%
\bibitem [{\citenamefont {Schnyder}\ \emph {et~al.}(2012)\citenamefont
  {Schnyder}, \citenamefont {Brydon},\ and\ \citenamefont {Timm}}]{Schnyder12}%
  \BibitemOpen
  \bibfield  {author} {\bibinfo {author} {\bibfnamefont {A.~P.}\ \bibnamefont
  {Schnyder}}, \bibinfo {author} {\bibfnamefont {P.~M.~R.}\ \bibnamefont
  {Brydon}}, \ and\ \bibinfo {author} {\bibfnamefont {C.}~\bibnamefont
  {Timm}},\ }\href {\doibase 10.1103/PhysRevB.85.024522} {\bibfield  {journal}
  {\bibinfo  {journal} {Phys. Rev. B}\ }\textbf {\bibinfo {volume} {85}},\
  \bibinfo {pages} {024522} (\bibinfo {year} {2012})}\BibitemShut {NoStop}%
\bibitem [{\citenamefont {Matsuura}\ \emph {et~al.}(2013)\citenamefont
  {Matsuura}, \citenamefont {Chang}, \citenamefont {Schnyder},\ and\
  \citenamefont {Ryu}}]{Matsuura13}%
  \BibitemOpen
  \bibfield  {author} {\bibinfo {author} {\bibfnamefont {S.}~\bibnamefont
  {Matsuura}}, \bibinfo {author} {\bibfnamefont {P.-Y.}\ \bibnamefont {Chang}},
  \bibinfo {author} {\bibfnamefont {A.~P.}\ \bibnamefont {Schnyder}}, \ and\
  \bibinfo {author} {\bibfnamefont {S.}~\bibnamefont {Ryu}},\ }\href
  {http://stacks.iop.org/1367-2630/15/i=6/a=065001} {\bibfield  {journal}
  {\bibinfo  {journal} {New Journal of Physics}\ }\textbf {\bibinfo {volume}
  {15}},\ \bibinfo {pages} {065001} (\bibinfo {year} {2013})}\BibitemShut
  {NoStop}%
\bibitem [{\citenamefont {Schnyder}\ and\ \citenamefont
  {Brydon}(2015)}]{Schnyder15}%
  \BibitemOpen
  \bibfield  {author} {\bibinfo {author} {\bibfnamefont {A.~P.}\ \bibnamefont
  {Schnyder}}\ and\ \bibinfo {author} {\bibfnamefont {P.~M.~R.}\ \bibnamefont
  {Brydon}},\ }\href {http://stacks.iop.org/0953-8984/27/i=24/a=243201}
  {\bibfield  {journal} {\bibinfo  {journal} {Journal of Physics: Condensed
  Matter}\ }\textbf {\bibinfo {volume} {27}},\ \bibinfo {pages} {243201}
  (\bibinfo {year} {2015})}\BibitemShut {NoStop}%
\bibitem [{\citenamefont {Sumita}\ and\ \citenamefont
  {Yanase}(2018)}]{Sumita18}%
  \BibitemOpen
  \bibfield  {author} {\bibinfo {author} {\bibfnamefont {S.}~\bibnamefont
  {Sumita}}\ and\ \bibinfo {author} {\bibfnamefont {Y.}~\bibnamefont
  {Yanase}},\ }\href {\doibase 10.1103/PhysRevB.97.134512} {\bibfield
  {journal} {\bibinfo  {journal} {Phys. Rev. B}\ }\textbf {\bibinfo {volume}
  {97}},\ \bibinfo {pages} {134512} (\bibinfo {year} {2018})}\BibitemShut
  {NoStop}%
\bibitem [{\citenamefont {Chung}\ and\ \citenamefont {Zhang}(2009)}]{Chung09}%
  \BibitemOpen
  \bibfield  {author} {\bibinfo {author} {\bibfnamefont {S.~B.}\ \bibnamefont
  {Chung}}\ and\ \bibinfo {author} {\bibfnamefont {S.-C.}\ \bibnamefont
  {Zhang}},\ }\href {\doibase 10.1103/PhysRevLett.103.235301} {\bibfield
  {journal} {\bibinfo  {journal} {Phys. Rev. Lett.}\ }\textbf {\bibinfo
  {volume} {103}},\ \bibinfo {pages} {235301} (\bibinfo {year}
  {2009})}\BibitemShut {NoStop}%
\bibitem [{\citenamefont {Fogelstr\"om}\ \emph {et~al.}(1997)\citenamefont
  {Fogelstr\"om}, \citenamefont {Rainer},\ and\ \citenamefont
  {Sauls}}]{Fogelstrom97}%
  \BibitemOpen
  \bibfield  {author} {\bibinfo {author} {\bibfnamefont {M.}~\bibnamefont
  {Fogelstr\"om}}, \bibinfo {author} {\bibfnamefont {D.}~\bibnamefont
  {Rainer}}, \ and\ \bibinfo {author} {\bibfnamefont {J.~A.}\ \bibnamefont
  {Sauls}},\ }\href {\doibase 10.1103/PhysRevLett.79.281} {\bibfield  {journal}
  {\bibinfo  {journal} {Phys. Rev. Lett.}\ }\textbf {\bibinfo {volume} {79}},\
  \bibinfo {pages} {281} (\bibinfo {year} {1997})}\BibitemShut {NoStop}%
\bibitem [{\citenamefont {Tanaka}\ and\ \citenamefont
  {Kashiwaya}(1995)}]{Tanaka95}%
  \BibitemOpen
  \bibfield  {author} {\bibinfo {author} {\bibfnamefont {Y.}~\bibnamefont
  {Tanaka}}\ and\ \bibinfo {author} {\bibfnamefont {S.}~\bibnamefont
  {Kashiwaya}},\ }\href {\doibase 10.1103/PhysRevLett.74.3451} {\bibfield
  {journal} {\bibinfo  {journal} {Phys. Rev. Lett.}\ }\textbf {\bibinfo
  {volume} {74}},\ \bibinfo {pages} {3451} (\bibinfo {year}
  {1995})}\BibitemShut {NoStop}%
\bibitem [{\citenamefont {Tanaka}\ \emph {et~al.}(2002)\citenamefont {Tanaka},
  \citenamefont {Tanuma}, \citenamefont {Kuroki},\ and\ \citenamefont
  {Kashiwaya}}]{Tanaka02}%
  \BibitemOpen
  \bibfield  {author} {\bibinfo {author} {\bibfnamefont {Y.}~\bibnamefont
  {Tanaka}}, \bibinfo {author} {\bibfnamefont {Y.}~\bibnamefont {Tanuma}},
  \bibinfo {author} {\bibfnamefont {K.}~\bibnamefont {Kuroki}}, \ and\ \bibinfo
  {author} {\bibfnamefont {S.}~\bibnamefont {Kashiwaya}},\ }\href {\doibase
  10.1143/JPSJ.71.2102} {\bibfield  {journal} {\bibinfo  {journal} {Journal of
  the Physical Society of Japan}\ }\textbf {\bibinfo {volume} {71}},\ \bibinfo
  {pages} {2102} (\bibinfo {year} {2002})}\BibitemShut {NoStop}%
\bibitem [{\citenamefont {Tanuma}\ \emph {et~al.}(2002)\citenamefont {Tanuma},
  \citenamefont {Kuroki}, \citenamefont {Tanaka}, \citenamefont {Arita},
  \citenamefont {Kashiwaya},\ and\ \citenamefont {Aoki}}]{Tanuma02}%
  \BibitemOpen
  \bibfield  {author} {\bibinfo {author} {\bibfnamefont {Y.}~\bibnamefont
  {Tanuma}}, \bibinfo {author} {\bibfnamefont {K.}~\bibnamefont {Kuroki}},
  \bibinfo {author} {\bibfnamefont {Y.}~\bibnamefont {Tanaka}}, \bibinfo
  {author} {\bibfnamefont {R.}~\bibnamefont {Arita}}, \bibinfo {author}
  {\bibfnamefont {S.}~\bibnamefont {Kashiwaya}}, \ and\ \bibinfo {author}
  {\bibfnamefont {H.}~\bibnamefont {Aoki}},\ }\href {\doibase
  10.1103/PhysRevB.66.094507} {\bibfield  {journal} {\bibinfo  {journal} {Phys.
  Rev. B}\ }\textbf {\bibinfo {volume} {66}},\ \bibinfo {pages} {094507}
  (\bibinfo {year} {2002})}\BibitemShut {NoStop}%
\bibitem [{\citenamefont {Tanaka}\ \emph {et~al.}(2009)\citenamefont {Tanaka},
  \citenamefont {Yokoyama}, \citenamefont {Balatsky},\ and\ \citenamefont
  {Nagaosa}}]{Tanaka09}%
  \BibitemOpen
  \bibfield  {author} {\bibinfo {author} {\bibfnamefont {Y.}~\bibnamefont
  {Tanaka}}, \bibinfo {author} {\bibfnamefont {T.}~\bibnamefont {Yokoyama}},
  \bibinfo {author} {\bibfnamefont {A.~V.}\ \bibnamefont {Balatsky}}, \ and\
  \bibinfo {author} {\bibfnamefont {N.}~\bibnamefont {Nagaosa}},\ }\href
  {\doibase 10.1103/PhysRevB.79.060505} {\bibfield  {journal} {\bibinfo
  {journal} {Phys. Rev. B}\ }\textbf {\bibinfo {volume} {79}},\ \bibinfo
  {pages} {060505} (\bibinfo {year} {2009})}\BibitemShut {NoStop}%
\bibitem [{\citenamefont {Tamura}\ \emph {et~al.}(2017)\citenamefont {Tamura},
  \citenamefont {Kobayashi}, \citenamefont {Bo},\ and\ \citenamefont
  {Tanaka}}]{Tamura17}%
  \BibitemOpen
  \bibfield  {author} {\bibinfo {author} {\bibfnamefont {S.}~\bibnamefont
  {Tamura}}, \bibinfo {author} {\bibfnamefont {S.}~\bibnamefont {Kobayashi}},
  \bibinfo {author} {\bibfnamefont {L.}~\bibnamefont {Bo}}, \ and\ \bibinfo
  {author} {\bibfnamefont {Y.}~\bibnamefont {Tanaka}},\ }\href {\doibase
  10.1103/PhysRevB.95.104511} {\bibfield  {journal} {\bibinfo  {journal} {Phys.
  Rev. B}\ }\textbf {\bibinfo {volume} {95}},\ \bibinfo {pages} {104511}
  (\bibinfo {year} {2017})}\BibitemShut {NoStop}%
\end{thebibliography}%

\newpage

\widetext
%\onecolumn

%%% Putting "S" to the Figure number
\renewcommand{\thefigure}{S\arabic{figure}} 

%%% Putting "S" to the Table number
\renewcommand{\thetable}{S\arabic{table}} 

%%% Putting "S" to the Section number
\renewcommand{\thesection}{S\arabic{section}.}

\renewcommand{\theequation}{S.\arabic{equation}}

\setcounter{figure}{0}
\setcounter{table}{0}
\setcounter{equation}{0}

%================================================= Supplementary Material
\begin{center} 
{\large {\bf Supplemental Materials: \\ Majorana multipole response of topological superconductors}}
\end{center}

%\baselineskip24pt

%================================

\begin{flushleft} 
{\bf S1.  Symmetry constraints on gap functions and $\mathcal{O}_{\Gamma}$'s}
\end{flushleft}

Here we discuss general symmetry constraints on helical Majorana fermions (MFs) appearing on a surface Brillouin zone (BZ) of time-reversal (TR) invariant topological superconductors (TSCs).
As relevant point group symmetry, we consider mirror reflections and rotations that are compatible to the surface.
We take into account all 2D point groups formed by them, 
$C_2, C_3, C_4, C_6, C_s, C_{2v}, C_{3v}, C_{4v}, C_{6v}$ in addition to TR symmetry.
These symmetries should be manifest in the Bogoliubov-de Gennes (BdG) Hamiltonian.
In the case of nematic superconductors, where a part of crystalline symmetry is spontaneously broken, we consider only the unbroken part. 

As is discussed in the main text, 
helical MFs show anisotropic behaviors under magnetic fields.
These behaviors are governed by two different magnetic point group symmetries, magnetic two-fold rotation and magnetic mirror reflection.  
In what follows, for the sake of concreteness, we consider the case where helical MFs  are protected by magnetic twofold rotation symmetry. The case where helical MFs are protected by magnetic mirror-reflection symmetry can be discussed in a similar fashion.

First, we consider symmetry constraints on gap functions. 
As pairing symmetry, we only need to take into account one-dimensional (1D) irreducible representations (IRs):
In general, if a gap function belongs to a higher dimensional IR, it spontaneously breaks crystalline symmetry and/or TR symmetry. 
In the former case, we should consider the unbroken part of crystalline symmetry, where the gap function belongs to a 1D IR. 
Furthermore, the latter case is excluded  since we consider TR-invariant TSCs.
%We 
% i.e. one is a one dimensional (1D) irreducible representation (IR) in terms of point group operations. 
Thus, we have the following constraint,
\begin{enumerate}
\item[(a)] The gap function is a 1D IR.
\end{enumerate} 
%
%This condition is generally imposed on gap functions because if the gap function does not belong to 1D IRs with  $C_{n}=\pm1$ and preserves TR symmetry, a node appears at helical MFs~\cite{Sumita18}. 
Additional constraints are required to obtain a nonzero $w_{\rm M1D}$. 
Consider the 1D winding number $w_{\rm M1D}$ with respect to two-fold rotation $C_2$,
\begin{eqnarray}
w_{M1D}=\frac{i}{4\pi}\int dk_{\perp}{\rm tr}\left[\Gamma_{\rm M} 
{\cal H}^{-1}(k_{\perp},{\bm k}_{\parallel})\partial_{k_\perp}{\cal H}(k_\perp,{\bm k}_{\parallel})\right], 
\end{eqnarray}
where ${\cal H}({\bm k})$ is the BdG Hamiltonian, $(k_{\perp}, {\bm k}_{\parallel})$ are momenta normal and parallel to the surface, and $\Gamma_{\rm M}\equiv e^{i\alpha}C_2{\cal T}{\cal C}$ is the magnetic chiral operator with respect to magnetic two-fold rotation.  
Here we choose $\alpha$ so as $\Gamma_{\rm M}^2=1$. 
When the gap function is even (odd) under two-fold rotation, the two-fold rotation operator  $C_2$ for the BdG Hamiltonian satisfies $[{\cal C}, C_2]=0$ ($\{{\cal C}, C_2\}=0$).
For $w_{\rm M1D}$ to be nonzero, however, $C_2$ should satisfy $[{\cal C}, C_2]=0$.
%$\Gamma_{\rm M}$ should satisfy $[\mathcal{C},\Gamma_{\rm M}]=[\mathcal{T},\Gamma_{\rm M}]=0$.
Actually, when $\{{\cal C}, C_2\}=0$, $\Gamma_{\rm M}$ with $\Gamma_{\rm M}^2=1$ is given as 
$\Gamma_{\rm M} = i C_2\mathcal{T}  \mathcal{C}$,  which leads to $\{\mathcal{T},\Gamma_{\rm M}\}=0$.
Then, when $\{\mathcal{T},\Gamma_{\rm M}\}=0$, the magnetic 1D winding number should be zero~\cite{Xiong17}.
%
% Thus, we require $[\mathcal{C},\Gamma_{\rm M}]=0$ so as to obtain a nonzero magnetic 1D winding number. In other words, since $[\mathcal{C},\Gamma_{\rm M}]=0$ reduces into $[\mathcal{C},C_2]=0$, we require the following constraint:
%This condition is verified by considering an opposite situation with $\{\mathcal{C},\Gamma_{\rm M}\}=0$. In such a case, we have 
%leading to 
%
Thus, we have the following constraint,
\begin{enumerate}
\item[(b)] The gap function is even under two-fold rotation.
\end{enumerate} 
Furthermore, when mirror-reflection symmetry $\sigma$ coexists, we obtain the following:
\begin{enumerate}
\item[(c)] The gap function is odd under mirror reflection.
\end{enumerate} 
This is because $w_{\rm M1D}=0$ if the gap function is even under mirror reflection.
When the gap function is even under reflection, it holds that $[\sigma,\mathcal{C}]=0$. 
Therefore, from $[\mathcal{T},\sigma]=\{C_2,\sigma\}=0$, we obtain $\{\sigma, \Gamma_{\rm M}\}=0$, which yields $w_{\rm M1D}=0$~\cite{Xiong17}. Using the conditions (a), (b), (c), and (b)' explained later, we can determine IRs of gap functions as shown in Table I in the main text. We note that if inversion symmetry $\mathcal{I}$ exists in the bulk superconductor, the magnetic 1D winding number is nonzero only when $\{\Gamma_{\rm M},\mathcal{I}\}=0$~\cite{Xiong17}, which leads to  $\{\mathcal{I}, \mathcal{C}\}=0$. The commutation relation implies that the gap function is odd under inversion which is consistent with the condition for obtaining nontrivial TSCs~\cite{Sato09,Sato10,Fu10}.

Next, we discuss symmetry constraints on $\mathcal{O}_{\Gamma}$'s. From the index theorem, a wave function of MFs at the zero energy $| u^{(a)}_0 \rangle$ satisfies
\begin{align}
 \Gamma_{\rm M }| u^{(a)}_0 \rangle = | u^{(a)}_0 \rangle,
\label{eq:chiral}
\end{align}
where $\Gamma_{ \rm M} =  C_2 \mathcal{T}\mathcal{C}$. Hence, we obtain $\Gamma_{\rm M}\rho^{(ab)}\Gamma_{\rm M}^{\dagger} =\rho^{(ab)} $, where $\rho^{(ab)} \equiv i(| u^{(a)}_0 \rangle \langle u^{(b)}_0 | -| u^{(b)}_0 \rangle \langle u^{(a)}_0 |)$. In order for the multipole response to exist, IRs of $\mathcal{O}_{\Gamma}$ need coincide with IRs of $\rho^{(ab)}$. So, $\mathcal{O}_{\Gamma}$ should satisfy
\begin{align}
 \Gamma_{\rm M} \mathcal{O}_{\Gamma} \Gamma_{\rm M} ^{\dagger} = \mathcal{O}_{\Gamma}.
\end{align}
From the definition, we have $\{\mathcal{C}, \mathcal{O}_{\Gamma}\}=0 $. 
Then, for magnetic responses, $\mathcal{O}_{\Gamma}$ should be odd under TR, i.e.$\{\mathcal{T}, \mathcal{O}_{\Gamma}\}=0$. 
Thus, we obtain
\begin{align}
 C_2\mathcal{O}_{\Gamma} C_2 ^{\dagger} = \mathcal{O}_{\Gamma},
\end{align}
which implies the following the constraint:
\begin{enumerate}
\item[(d)] $\mathcal{O}_{\Gamma}$ is a 1D IR with $C_2=1$.
\end{enumerate} 

Now, we take into account mirror-reflection symmetry.
The constraint (c) implies that $\{{\cal C}, \sigma\}=0$, and by combining it with 
$
[\mathcal{T},\sigma]=\{C_2, \sigma\}=0, 
$
we obtain $[\Gamma_{\rm M},\sigma]=0$.
Therefore, MFs at the zero energy $|u_0^{(a)}\rangle$ can be simultaneous eigenstates of $\Gamma_{\rm M}$ and $\sigma$. 
For a Kramers pair of MFs $|u^{(a)}_0 \rangle$ ($a=1,2$), we place the relations
\begin{align}
 &\mathcal{C}|u^{(a)}_0 \rangle  = |u^{(a)}_0 \rangle, \ \ (a=1,2), \label{eq:PH}  \\
 &\mathcal{T}|u^{(1)}_0 \rangle  = |u^{(2)}_0 \rangle,  \label{eq:TR}
\end{align}
without loss of generality \footnote{These relations lead to $\{\mathcal{C}, \rho^{(12)}\}=\{\mathcal{T}, \rho^{(12)}\}=0$, which are consistent with $\{{\cal C}, {\cal O}_{\Gamma}\}=\{{\cal T}, {\cal O}_{\Gamma}\}=0$.}.
Then, from $[{\cal T}, \sigma]=0$, we have
\begin{align}
\sigma |u^{(a)}_0 \rangle = (-)^{a}i|u^{(a)}_0 \rangle, \ \ (a=1,2). \label{eq:sigma-basis}
\end{align}
in addition to Eq.(\ref{eq:chiral}).
Equation~(\ref{eq:sigma-basis}) leads to $\sigma \rho^{(12)} \sigma^{-1} =- \rho^{(12)}$, implying that 
\begin{align}
 \sigma \mathcal{O}_{\Gamma} \sigma^{-1} = -\mathcal{O}_{\Gamma}.
\label{eq:Osigma}
\end{align}
Thus, we find the following constraint:
\begin{enumerate}
\item[(e)] $\mathcal{O}_{\Gamma}$ is a 1D IR with $\sigma=-1$.
\end{enumerate} 
When there are multiple Kramers pairs of MFs, $\rho^{(ab)}$ can contain other IRs which behave differently from Eq.(\ref{eq:Osigma}). But we find that the other IRs only gives a subleading contribution for magnetic responses.

Finally, we discuss the action of $C_{n}$ ($n \ge3$)  on  ${\cal O}_{\Gamma} $. 
From the constraint (a), $C_n$ may have two possible commutation relations with ${\cal C}$:
When the gap function is even (odd) under $C_n$, then we have $[C_n, {\cal C}]=0$ ($\{C_n, {\cal C}\}=0$).
However, stable helical MFs protected by magnetic two fold rotation are possible only when $[C_n, {\cal C}]=0$:
When $\{C_n, {\cal C}\}=0$, it holds that $\{\Gamma_{\rm M}, C_n\}=0$, and thus  
$C_n|u_0^{(a)}\rangle$ and $|u_0^{(a)}\rangle$ have opposite chirailties with respect to $\Gamma_{\rm M}$. As a result, they can be easily gapped out without topological protection.
Therefore, we have the following constraint
\begin{enumerate}
\item[(b)'] The gap function is even under $C_n$,
\end{enumerate} 
which includes the constraint (b) as a special case.
When the constraint (b)' is satisfied, we have $[{\cal C}, C_n]=0$, and thus from $[C_n, {\cal T}]=[C_2, C_n]=0$, we can show that $[\Gamma_{\rm M}, C_n]=0$.
%\begin{align}
%[\mathcal{C},C_n]=[\mathcal{T},C_n]=[C_2, C_n]=0. \label{eq:cn-comm}
%\end{align}
Therefore, 
$|u^{(a)}_0 \rangle$ is a simultaneous eigenstate of $\Gamma_{\rm M}$ and $C_n$, 
 \begin{align}
C_n |u^{(a)}_0 \rangle = \lambda^{(a)}|u^{(a)}_0 \rangle, \label{eq:cn-basis}
\end{align}
where $\lambda^{(a)}$ is an eigenvalue of $C_n$. 
For a Kramer pair of MFs $|u_n^{(a)}\rangle$ ($a=1,2$), we also have Eqs.~(\ref{eq:PH}) and (\ref{eq:TR}), which lead to $\lambda^{(1)}=\lambda^{(2)}=1$.
%Since
%\begin{align}
% C_n |u^{(a)}_0 \rangle = C_n\mathcal{C} |u^{(a)}_0 \rangle = \mathcal{C} C_n |u^{(a)}_0 \rangl%e  =\mathcal{C} \lambda^{(a)} |u^{(a)}_0 \rangle =(\lambda^{(a)} )^{\ast}\mathcal{C} |u^{(a)}_0% \rangle =(\lambda^{(a)})^{\ast} |u^{(a)}_0 \rangle,
%\end{align}
%Thus,  $\lambda^{(a)} =(\lambda^{(a)})^{\ast} $, reading $\lambda^{(a)} = \pm 1$.  Also, TR symmetry imposes $\lambda^{(1)} =\lambda^{(2)}$. 
Therefore, we obtain $C_n \rho^{(12)} C_n^{-1} = \rho^{(12)}$, implying that
\begin{align}
 C_n \mathcal{O}_{\Gamma} C_n^{\dagger} = \mathcal{O}_{\Gamma}.
\label{eq:Ocn}
\end{align}
Thus, we find the following constraint:
\begin{enumerate}
\item[(d)'] $\mathcal{O}_{\Gamma}$ is a 1D IR with $C_n=1$,
\end{enumerate} 
which include (d) as a special case.
Again, when there are multiple Kramers pairs of MFs, $\rho^{(ab)}$ can contain other IRs which behave differently from Eq.(\ref{eq:Ocn}) with $n\ge 3$, but they only gives a subleading contribution for magnetic responses.

Using the constraints (d)' and (e), we can determine IRs of $\mathcal{O}_{\Gamma}$'s with the leading contributions in Table I in the main text. Note that subleading contributions may exist when there are more than two Kramers pairs of MFs,  which do not necessarily satisfy the constraints (e) and (d)' with $n\ge 3$.

\begin{flushleft} 
{\bf S2. Magnetic response of superconducting topological insulators }
\end{flushleft} 

As is shown in the main text, helical MFs in spin-3/2 superconductors show the magnetic octupole response on a surface preserving $C_{\rm 3v}$ symmetry. 
Here we examine magnetic responses of MFs in spin-1/2 superconductors on a surface with the same symmetry.
In contrast to the spin-3/2 case, we find that MFs in spin-1/2 superconductors do not show the magnetic octupole response in the leading order.

Here we consider the superconducting doped topological insulator (TI), {\it A}$_{x}$Bi$_2$Se$_3$ ($A$=Cu, Sr, Nb), as a representative example of spin-1/2 superconductors.
The crystal point group is $D_{3d}$, and thus a surface perpendicular to the $c$-axis hosts $C_{\rm 3v}$ symmetry.
The system consists of two bands near the Fermi energy, which are predominated by Se $p_z$ orbitals on the top and bottom layer of the unit cell.
These orbital degrees of freedom do not provide any non-trivial contribution under rotation around the $c$-axis, and thus these bands behave as ordinary spin-1/2 electrons under $C_{3v}$.

The low-energy Hamiltonian in the normal state is given by~\cite{Fu10}
\begin{align}
H_{\rm TI}(\bm{k}) = c(\bm{k}) + m(\bm{k}) \sigma_x + v_z k_z \sigma_y + v (k_x s_y - k_y s_x) \sigma_z + \lambda (k_x^3 - 3 k_x k_y^2) s_z, \label{eq:CuxBi2Se3}
\end{align}
with $c(\bm{k}) =c_0+ c_1 k_z^2 + c_1 (k_x^2+k_y^2)$ and $m(\bm{k}) =m_0+ m_1 k_z^2 + m_1 (k_x^2+k_y^2)$. 
Here $\sigma_i$
and $s_i$ are the Pauli matrices in the orbital and spin spaces, respectively. 
The last term proportional to $\lambda$ is the hexagonal warping term. The symmetries in Eq.~(\ref{eq:CuxBi2Se3}) are TR symmetry $T = i s_y K$, inversion symmetry $I=\sigma_x$, three-fold rotation around the $c$ axis $C_3 = e^{-i \frac{\pi}{3} s_z} $, and the vertical mirror reflection $\sigma_v (yz) = i s_x$ with respect to the $yz$ plane. 
Importantly, at the $k_x=k_y=0$, where Eq.~(\ref{eq:CuxBi2Se3}) is reduced to $H_{\rm TI} (0,0,k_z) = c(0,0,k_z) + m(0,0,k_z) \sigma_z$,  the three-fold rotation symmetry becomes fully rotational invariance with $C_{\infty} = e^{i \frac{s_z}{2} \theta}$ ($0 \le \theta < 2 \pi$).
As we shall explained in S3, this symmetry enhancement is intrinsic to spin-1/2 systems.

In superconducting states, the BdG Hamiltonian is given by 
\begin{align}
 \mathcal{H}_{\rm TI}(\bm{k}) = \begin{pmatrix} H_{\rm TI } (\bm{k}) -\mu & \Delta(\bm{k}) \\ \Delta(\bm{k})^{\dagger} & -H_{\rm TI}(-\bm{k})^t + \mu  \end{pmatrix},
\end{align}
where $\mu$ is the chemical potential and $\Delta(\bm{k})$ is the gap function. Due to the Fermi statistics, we have six onsite gap functions: $(\Delta_0 i s_y, \Delta_0 i \sigma_x s_y, \Delta_0 i \sigma_z s_y, \Delta_0 \sigma_y s_z, \Delta_0 i \sigma_y, \Delta_0  \sigma_y s_x)$, which are classified within $D_{3d}$ such that $(A_{1g}, A_{1g}, A_{2u}, E_{u}, E_{u}, A_{1u})$. It has been known that the $A_{1u}$ gap function realizes full-gap TSC~\cite{Fu10} and the $E_u$ gap function the nematic superconductor~\cite{Fu14,Matano16, Yonezawa16}. Here we examine magnetic responses for the $A_{2u}$ and $A_{1u}$ gap functions. 

\begin{figure}[tbp]
\centering
 \includegraphics[width=16cm]{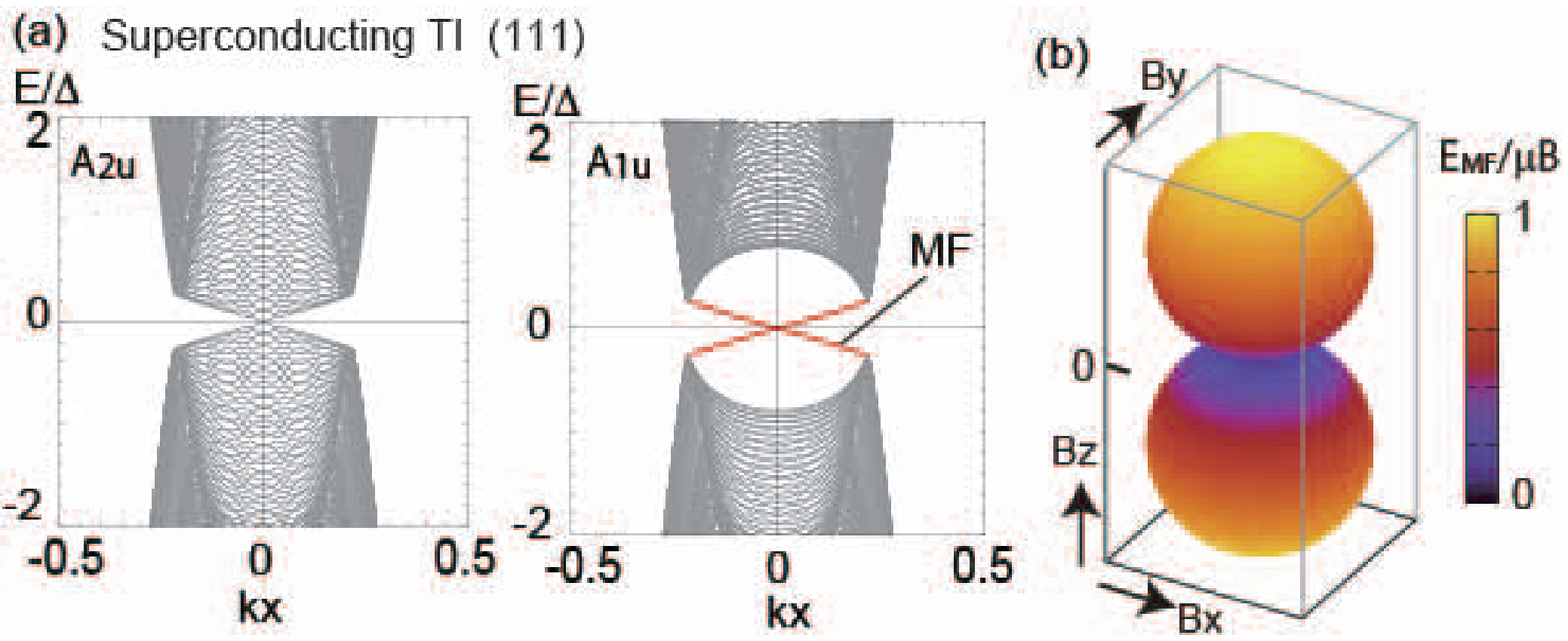}
 \caption{(a) Surface energy spectrum of Eq.~(\ref{eq:CuxBi2Se3}) at the (111) plane, where we replace $k_i$, $k_i^2$ with $\sin k_i$, $2(1-\cos k_i)$ and the (111) plane is perpendicular to $c$ axis. $(c_0, c_1, c_2, m_0 , m_1 , m_2, v_z ,v, \mu, \Delta)=(-0.0083, 5.74, 30.4, -0.28, 6.84, 44.5, 3.33, 2.26, 3, 1)$ (b) Energy gap of the helical MF at $k_x = k_y=0$ in the superconducting TI with the $A_{1u}$ gap function.}\label{fig:STI}
\end{figure}

First, we consider the $A_{2u}$ gap function, which belongs to $A_1$ in $C_{3v}$ at the surface BZ. 
Although the spin-3/2 case with the same pairing symmetry hosts MFs with a magnetic octupole response, the present case do not.  As is shown in Fig.~\ref{fig:STI} (a), there appear  point nodes on the $C_{ 3v}$ symmetric line, so no clear MFs is obtained.
This is due to the spin-1/2 nature of the system.
As $C_{3}$ is enhanced to $C_\infty$ on the $k_z$-axis, the $A_{2u}$ pairing symmetry cannot support a nonzero gap on the $k_z$-axis.
(See also the group theoretical analysis in  \cite{Sumita18}.)
%Thus, in order to realize the octupole response, high spin TSCs with spin $3/2$ fermions are required.

In the case of the $A_{1u}$ gap function, we have a helical MF on a surface normal to the $c$-axis.
See Fig.~\ref{fig:STI} (a). Applying the magnetic Zeeman field $\mu \bm{B} \cdot \bm{s}$, we calculate the energy gap of the helical MF. 
As illustrated in Fig.~\ref{fig:STI} (b), 
the energy gap results in the dipole response with respect to the $c$-axis.
Again, this behavior is due to the spin-1/2 nature of the system: 
Because of the enhanced rotation symmetry $C_{\infty}$, we can define the magnetic 1D winding number by using $C_2$ subgroup of $C_\infty$.
The magnetic winding number is nonzero, and thus the helical MF remains gapless as long as one keeps the magnetic symmetry of $C_2$.
The magnetic symmetry is broken under a magnetic field with a nonzero component along the $c$-axis, so we have the magnetic dipole response.

\begin{flushleft} 
{\bf S3. Enhancement of rotational symmetry in spin-1/2 systems}
\end{flushleft} 
 
We discuss here that the enhancement of rotational symmetry is specific to spin-1/2 systems. We start from an effective Hamiltonian with basis $|\pm j_z \rangle$, which is generally described by 
\begin{align}
H(\bm{k}) = a_0(\bm{k}) s_0 + a_x(\bm{k}) s_x  + a_y(\bm{k}) s_y +  a_z(\bm{k}) s_z,  \label{eq:generalHami}
\end{align}
where $s_i$ are the Pauli matrices with basis $|\pm j_z \rangle$ and  $a_i (\bm{k})$ are real functions of $\bm{k}$. We are interested in the behavior of the Hamiltonian on a high symmetric axis where the enhancement of rotational symmetry may occur. Hereafter, we focus on a $C_{\rm 3v}$ symmetric axis, say $k_z$ axis. Practically, there exists a $C_{\rm 3v}$ symmetric axis along the $[111]$ direction in the superconducting TIs and half-Heusler superconductors. On the $k_z$ line, the Hamiltonian (\ref{eq:generalHami}) need to satisfy
\begin{subequations}
\label{eq:C3vHami}
\begin{align}
&C_{3}^{\dagger} H(0,0,k_z) C_3 = H(0,0,k_z), \\
& {\sigma_v}^{\dagger} H(0,0,k_z) \sigma_v = H(0,0,k_z),
\end{align} 
\end{subequations}
where $C_3$ and $\sigma_v$ are given by
\begin{subequations}
\label{eq:C3vop}
\begin{align}
 &C_{3} = \diag (e^{-i \frac{2\pi}{3}j_z },e^{i \frac{2\pi}{3} j_z}), \\
 &\sigma_v = i s_x,
\end{align}
\end{subequations}
%where $s_i$ are the Pauli matrices with basis $|\pm j_z \rangle$. To keep the $C_{3v}$ invariance, an effective Hamiltonian need to satisfy
From Eqs.~(\ref{eq:generalHami}) and (\ref{eq:C3vHami}), the enhancement of rotation symmetry occurs only if the symmetry constraints~(\ref{eq:C3vHami}) demand $a_x=a_y=0$. In such a case, we achieve $[H(0,0,k_z),s_z]=0$, so the Hamiltonian hosts $C_{\infty {\rm v}}$ symmetry on the $k_z$ axis, where $C_3$ is extended to $C_{\infty}$ such that 
 \begin{align}
   C_{\infty}^{\dagger} H(0,0,k_z) C_{\infty} = H(0,0,k_z), 
\end{align} 
with $C_{\infty} = \diag (e^{-i \theta j_z },e^{i \theta j_z})$ ($0 \le \theta < 2 \pi$).  Note that the similar argument is applicable to the case without $\sigma_v$. %, in which the $C_3$ symmetry is extended to the $C_{\infty}$ symmetry.  
In the following, we consider the symmetry constraints for $j_z=1/2$ and $3/2$.

%  A symmetry adopted Hamiltonian satisfying Eqs.~(\ref{eq:C3vHami}) and show that $j_z=3/2$ allows non-vanishing terms proportional to $s_x, s_y$ but $j_z=1/2$ does not. The absence of terms proportional to $s_x, s_y$ reads the existence of an additional symmetry, $[H(0,0,k_z),s_z]=0$. Thus, the Hamiltonian satisfies a full-rotational invariant
%
 When $j_z = 1/2$, Eqs.~(\ref{eq:C3vop}) are recast into
\begin{subequations}
\label{eq:C3vop-onehalf}
\begin{align}
 &C_{3} = e^{-i \frac{\pi}{3} s_z } \\
 &\sigma_v = i s_x.
\end{align}
\end{subequations}
Using Eqs.~(\ref{eq:generalHami}) and (\ref{eq:C3vop-onehalf}), we readily find that Eqs.~(\ref{eq:C3vHami})
are satisfied only when $a_x=a_y=a_z=0$; namely, the enhancement is inevitable for spin-1/2.

On the other hand, when $j_z = 3/2$, Eqs.~(\ref{eq:C3vop}) are recast into
\begin{subequations}
\label{eq:C3vop-threehalf}
\begin{align}
 &C_{3} = - s_0 \\
 &\sigma_v = i s_x.
\end{align}
\end{subequations}
Using Eqs.~(\ref{eq:generalHami}) and (\ref{eq:C3vop-threehalf}), an effective Hamiltonian satisfying Eqs.~(\ref{eq:C3vHami}) is constructed as
\begin{align}
H(0,0,k_z) = a_0(k_z) s_0 + b k_z s_x ,  
\end{align}
where $b$ is a real coefficient. Therefore, the enhancement does not occur for spin-3/2.

\begin{flushleft} 
{\bf S4. Two-orbital model with magnetic octupole response }
\end{flushleft} 

%In Table I in the main manuscript, we find the magnetic octupole  response in systems with $C_{3v}$ and $C_{6v}$ point group symmetry. 
Here we present a two-orbital model with magnetic octupole response.
%Here, using a toy model that hosts the magnetic octupole response, we calculate $\hat{\mathcal{O}}_{\rm MF} = \sum_{ab} \gamma^{(a)} \gamma^{(b)} \langle u_0^{(a)} | \mathcal{O} | u_{0}^{(b)} \rangle$ with $\mathcal{O} = \diag (O,-O^T)$ explicitly and prove that the Majorana octupole appears as a leading contribution. To this end, 
Being different from the two-orbital system in S2, the present system consists of $p_x$ and $p_y$-orbitals,
which host $l_z=\pm 1$ angular momenta in the $z$-direction.
Therefore, it contains high spin electrons with $|J_z|=3/2$.
The Hamiltonian in the normal state is given by
\begin{align}
H({\bm k})=C({\bm k})\sigma_0s_0+\lambda_1\sigma_ys_z+\lambda_2\sin k_z(\sigma_xs_x+\sigma_zs_y), 
\end{align}
where $C(\bm{k}) = c_0+t_{xy}\Big\{\cos k_x +\cos \left( -\frac{1}{2}k_x+\frac{\sqrt{3}}{2}k_y\right)+\cos \left( -\frac{1}{2}k_x-\frac{\sqrt{3}}{2}k_y\right)\Big\} + t_z \cos k_z $,
and $\lambda_i$ $(i=1,2)$ are the spin-orbit couplings.
Here, $\sigma_i$ and $s_i$ describe the Pauli matrices in orbital and spin spaces.
$\sigma_z=1$ ($\sigma_z=-1$) represents the $p_x$-orbital ($p_y$-orbital).
We assume $D_{3h}$ symmetry, which is generated by
\begin{align}
 &\sigma_v = -i \sigma_z s_y, \label{eq:MRop} \\
&\sigma_h = -i \sigma_0 s_z, \label{eq:MRop} \\
 &C_3 = \begin{pmatrix} - \frac{1}{2} & - \frac{\sqrt{3}}{2} \\ \frac{\sqrt{3}}{2} & - \frac{1}{2}\end{pmatrix}_{\sigma} \otimes e^{-i \frac{\pi}{3} s_z}, \label{eq:C3op}
\end{align}
where $\sigma_v$ ($\sigma_h$) is the mirror-reflection operators with respect to the $zx$ ($xy$) plane, and $C_3$ the three-fold rotation operator around the $k_z$-axis.

We consider the superconducting state which is described by the BdG Hamiltonian 
\begin{align}
\mathcal{H}({\bm k})=
\left(
\begin{array}{cc}
H({\bm k})-\mu & \Delta({\bm k})\\
\Delta({\bm k})^{\dagger}& -H({\bm k})^t+\mu     
\end{array}
\right), 
\end{align}
with the gap function $\Delta({\bm k})$ 
\begin{align}
\Delta({\bm k})=i\Delta_0s_y+\Delta_1\sigma_ys_x+\Delta_2\sin k_z(-\sigma_xs_z+is_0\sigma_z). 
\end{align}
%consisting of $p_x$ and $p_y$ orbitals,
%\begin{align}
% \tilde{H}(\bm{k}) =& -\mu \tau_z + C(\bm{k})  \tau_z + \lambda_1 \sigma_y s_z  + \lambda_2 \sin k_z (\sigma_x s_x  +\sigma_z s_y \tau_z) \notag \\ 
%&-\Delta_0 s_y \tau_y + \Delta_1 \sigma_y s_x \tau_x + \Delta_2 \sin k_z (-\sigma_x s_z \tau_x - \sigma_z  \tau_y), 
%\end{align}  
Here $\Delta_0$, $\Delta_1$, and $\Delta_2$ are constants, and the gap function belongs to the $A_1'$ IR of $D_{3h}$.  
%$\mu$ is the chemical potential, 
 The model has three bands described by $E_1(\bm{k}) = C(\bm{k}) - \lambda_1$, $E_2(\bm{k}) = C(\bm{k}) + \lambda_1 + 2 \lambda_2 \sin k_z$, and $E_3(\bm{k}) = C(\bm{k}) + \lambda_1 - 2 \lambda_2 \sin k_z $, where only the $E_1$ band is doubly degenerate.
Below we consider the case where $\Delta_2$ is dominant and the chemical potential $\mu$ lies on $E_2$ and $E_3$ bands.
In this case, the ($001$) surface hosts a helical MF with flat band  associated with the line node at $k_z=0$. The $A_1'$ pairing symmetry becomes  $A_1$ for the surface point group $C_{3v}$.
For the BdG Hamiltonian, the $D_{3h}$ point group operations are given by
\begin{align}
&\tilde{\sigma}_v=\diag(\sigma_v, \sigma_v^*)=-i\sigma_x s_y \tau_0,\notag\\
&\tilde{\sigma}_h= \diag(\sigma_h, \sigma_h^*)=-i\sigma_0 s_z \tau_z, \notag\\
&\tilde{C}_3=\diag(C_3, C_3^*),
\end{align} 
where $\tau_\mu$ are the Pauli matrices in the Nambu space.

\begin{table*}[tbp]
\centering
\caption{Possible matrices $O=s_\mu\sigma_\nu$ for local operators of the two-orbital model in S4.
We classify $O$ by IRs of $D_{3h}$.} 
\label{tab:SIR-D3h}
\begin{tabular}{cccc}
\hline \hline
IR of $D_{3h}$ &TR-even&TR-odd& basis by magnetic fields \\ \hline
 $A'_1$& $\sigma_0 s_0, \sigma_y s_z$ &   & \\
 $A'_2$& & $\sigma_0s_z, \sigma_y s_0$  &$B_z$ \\
 $E'$& $\{\sigma_x s_0,\sigma_z s_0\}$&$\{\sigma_x s_z, \sigma_z s_z\}$ & \\
 $A_1''$& & $\sigma_z s_x -\sigma_x s_y$ & $B_x^3-3B_xB_y^2$\\
 $A_2''$& & $\sigma_z s_y +\sigma_x s_x$ & $B_y^3-3B_yB_x^2$ \\
 $E''$ & $\{\sigma_y s_x, \sigma_y s_y\}$& $\{\sigma_0 s_x, \sigma_0 s_y\},\{\sigma_z s_x +\sigma_x s_y, \sigma_x s_x -\sigma_z s_y \}$& $\{B_x,B_y\}$ \\
\hline \hline
\end{tabular}
\end{table*}

In the following, we examine magnetic responses of the helical MF.
In the above situation, the following magnetic 1D winding number becomes 2. 
 \begin{align}
W(\Gamma_{\sigma}) = \frac{i}{4\pi} \int^{\pi}_{-\pi} d k_z\; \trace \left[ \Gamma_{\sigma} \mathcal{H}(\bm{k})^{-1} \partial_{k_z} \mathcal{H}(\bm{k})\right] |_{k_x=k_y=0}, \label{eq:1dwinding-toy}
\end{align}
with $\Gamma_{\sigma} = \tilde{\sigma}_v \mathcal{T} \mathcal{C} = s_0\sigma_z  \tau_x$.
%$\mathcal{C} = \tau_x K$ and $\mathcal{T}=i s_yK$, respectively, where $K$ is the complex conjugation. 
The corresponding (Kramers) pair of zero modes $|u_0^{(a)}\rangle$ ($a=1,2$) form a helical MF.  
From the index theorem, the zero modes satisfy $\Gamma_\sigma|u_0^{(a)}\rangle=|u_0^{(a)}\rangle$, and thus they are given by
  \begin{align}
   | u_0^{(1)} \rangle 
= \left(
\begin{array}{c}
\left(
\begin{array}{c}
\,\,\alpha \,\,\\
\beta
\end{array}
\right)_\sigma
\\[5mm]
\left(
\begin{array}{c}
\alpha \\
-\beta
\end{array}
\right)_\sigma
\end{array} 
\right)_{\tau} \otimes u_s, \ \   | u_0^{(2)} \rangle =  \mathcal{T} | u_0^{(1)} \rangle, \label{eq:state_of_sigmav}
 \end{align} 
where $\alpha$ and $\beta$ are real coefficients and $u_s$ is an arbitrary function in the spin space. 
We also have a constraint from $C_3$ symmetry:
From $C_3$ symmetry, we obtain two additional magnetic chiral operators, $\tilde{C}^{-1}_3\Gamma_\sigma \tilde{C}_3$ and $\tilde{C}^{-2}_3\Gamma_\sigma \tilde{C}^2_3$.
They define two additional 1D winding numbers, $W(\tilde{C}_3^{-1}\Gamma_{\sigma_v}\tilde{C}_3)$ and $W(\tilde{C}_3^{-2}\Gamma_{\sigma_v}\tilde{C}_3^2)$, which are equal to $W(\Gamma_{\sigma})=2$. 
The index theorem tells us again that
$\tilde{C}_3^{-1} \Gamma_{\sigma}\tilde{C}_3|u^{(a)}_0 \rangle =|u_0^{(a)}\rangle$ and 
$\tilde{C}_3^{-2}\Gamma_{\sigma} \tilde{C}_3^2|u^{(a)}_0 \rangle=|u^{(a)}_0 \rangle $, which are recast into $\Gamma_\sigma \tilde{C}_3|u_0^{(a)}\rangle=\tilde{C}_3|u_0^{(a)}\rangle$ and $\Gamma_\sigma \tilde{C}^2_3|u_0^{(a)}\rangle=\tilde{C}^2_3|u_0^{(a)}\rangle$.
In other words, $|u_0^{(a)}\rangle$, $\tilde{C}_3|u_0^{(a)}\rangle$ and $\tilde{C}_3^2|u_0^{(a)}\rangle$ are eigenstates of $\Gamma_\sigma$ with the same eigenvalue $\Gamma_\sigma=1$.
Moreover, we can obtain an eigenstate of $\tilde{C}_3$ by combining these states. 
Therefore, the zero modes can be simultaneous eigenstates of $\Gamma_\sigma$ and $\tilde{C}_3$.
Since $\Gamma_\sigma$ and $\tilde{C}_3$ obey $\Gamma_\sigma \tilde{C}_3 \Gamma_\sigma=\tilde{C}_3^{-1}$, the eigenvalue of $\tilde{C}_3$ must be $-1$.
%
%Since the gap function belongs to $A_1$ for the surface point group $C_{3v}$, the $C_3$ operation for the BdG Hamiltonian commutes with ${\cal C}$,  
%At the $C_{3v}$ symmetric point, the BdG Hamiltonian satisfies $C_3^{\dagger} H(0) C_3 = H(0)$ and $\Gamma_{\sigma}^{\dagger} H(0) \Gamma_{\sigma} =-H(0)$ simultaneously. Thus, the magnetic 1D winding number satisfies $W(\Gamma_{\sigma_v}) = 
%W(C_3^{\dagger}\Gamma_{\sigma_v}C_3)=W((C_3^{\dagger})^2\Gamma_{\sigma_v}(C_3)^2)$. 
%
%So, from the index theorem, zero modes must satisfy 
% implying that $\sigma_v$ commutes with $C_3$ and $|u^{(a)}_0 \rangle $ is a simultaneous eigenstate of $\Gamma_{\sigma}$ and $C_3$. 
Imposing the eigenvalue condition for $\tilde{C}_3$ on Eq.~(\ref{eq:state_of_sigmav}), we obtain
\begin{align}
     | u_0^{(1)} \rangle =  \left( \begin{array}{@{\,} c @{\,}}C\\ i C \\ D \\ -iD \\ C\\ -iC \\ D\\ iD\end{array} \right) , \ \ | u_0^{(2)} \rangle =  \mathcal{T} | u_0^{(1)} \rangle,
\label{eq:state_of_c3} 
\end{align} 
where $C$ and $D$ are real coefficients, and we take the Nambu space as 
$(\hat{c}_{1\uparrow}, \hat{c}_{1\downarrow}, \hat{c}_{2\uparrow}, \hat{c}_{2\downarrow}, 
\hat{c}^{\dagger}_{1\uparrow}, \hat{c}^{\dagger}_{1\downarrow}, \hat{c}^{\dagger}_{2\uparrow}, \hat{c}^{\dagger}_{2\downarrow} 
)^t$ with the spin $s=(\uparrow, \downarrow)$ and the orbital $\sigma=(1,2)$.

Now we perform the mode expansion of the quantum field
\begin{align}
    \bm{\Psi}(x) = \left( \begin{array}{@{\,} c @{\,}}\hat{c}_{1s} (x) \\ \hat{c}_{2s} (x) \\ \hat{c}_{1s}(x)^{\dagger} \\ \hat{c}_{2s} (x)^{\dagger} \end{array} \right) =  \sum_{a=1,2}  | u_0^{(a)} \rangle  \hat{\gamma}^{(a)} + (\text{non-zero energy mode}).
\end{align}
Neglecting the non-zero energy modes, we find 
\begin{align}
&\hat{c}_{1s} = \hat{c}_{1s}^{\dagger}, \quad \hat{c}_{2s} = -\hat{c}_{2s}^{\dagger}.\\
&\hat{c}_{2 \uparrow} = i\hat{c}_{1 \uparrow} =- \hat{c}_{2 \uparrow}^{\dagger}=i \hat{c}_{1 \uparrow}^{\dagger}, \ \ \hat{c}_{2 \downarrow} = -i\hat{c}_{1 \downarrow} =- \hat{c}_{2 \downarrow}^{\dagger}=-i \hat{c}_{1 \downarrow}^{\dagger}. 
\end{align}
from Eqs.(\ref{eq:state_of_sigmav}) and (\ref{eq:state_of_c3}).

Using these relations, we find that the local density and spin density operators of the zero modes vanish such that
\begin{align}
 \hat{\rho}_{\rm MF} &=\frac{1}{2}\bm{\Psi}(x)^{\dagger} \diag(s_0\sigma_0,-s_0\sigma_0) \bm{\Psi}(x)|_{\rm MF} \notag \\ &= \frac{1}{2} \left( c_{1s}^{\dagger} c_{1s} + c_{2s}^{\dagger} c_{2s} - c_{1s} c_{1s}^{\dagger}-c_{2s} c_{2s}^{\dagger} \right) \notag \\
                   &=  \frac{1}{2} \left( c_{1s}^{\dagger} c_{1s}^{\dagger} - c_{2s}^{\dagger} c_{2s}^{\dagger} - c_{1s}^{\dagger} c_{1s}^{\dagger}+c_{2s}^{\dagger} c_{2s}^{\dagger} \right) \notag \\
                   &= 0. \\
 (\hat{S}_{i=x,z})_{\rm MF} &=\frac{1}{4}\bm{\Psi}(x)^{\dagger} \diag(s_i\sigma_0,-s_i\sigma_0) \bm{\Psi}(x)|_{\rm MF} \notag \\ 
     &= \frac{1}{4} ( c_{1s}^{\dagger} (s_i)_{ss'}c_{1s'} + c_{2s}^{\dagger} (s_i)_{ss'}c_{2s'} - c_{1s} (s_i^t)_{ss'}c_{1s'}^{\dagger}-c_{2s} (s_i^t)_{ss'} c_{2s'}^{\dagger} ) \notag \\
                   &=  \frac{1}{4} (c_{1s}^{\dagger} (s_i)_{ss'}c_{1s'}^{\dagger} - c_{2s}^{\dagger} (s_i)_{ss'}c_{2s'}^{\dagger} - c_{1s}^{\dagger} (s_i^t)_{ss'}c_{1s'}^{\dagger}+c_{2s}^{\dagger} (s_i^t)_{ss'} c_{2s'}^{\dagger} )\notag\\
&=0.\notag\\
(\hat{S}_y)_{\rm MF} &=\frac{1}{4}\bm{\Psi}(x)^{\dagger} \diag(s_y\sigma_0,s_y\sigma_0) \bm{\Psi}(x)|_{\rm MF} \notag \\
&= \frac{1}{2} \left(\hat{c}_{1s}^{\dagger} (s_y)_{ss'}\hat{c}_{1s'}^{\dagger}-\hat{c}_{2s}^{\dagger} ( s_y )_{ss'}\hat{c}_{2s'}^{\dagger} \right) \notag \\
    &=\frac{1}{2} \left(-i\hat{c}_{1\uparrow}^{\dagger}\hat{c}_{1\downarrow}^{\dagger}+i\hat{c}_{1\downarrow}^{\dagger}\hat{c}_{1\uparrow}^{\dagger}+i\hat{c}_{2\uparrow}^{\dagger}\hat{c}_{2\downarrow}^{\dagger} -i\hat{c}_{2\downarrow}^{\dagger}\hat{c}_{2\uparrow}^{\dagger} \right) \notag \\
    &=\frac{1}{2} \left(-i\hat{c}_{1\uparrow}^{\dagger}\hat{c}_{1\downarrow}^{\dagger}+i\hat{c}_{1\downarrow}^{\dagger}\hat{c}_{1\uparrow}^{\dagger}+i(-i)\hat{c}_{1\uparrow}^{\dagger}i\hat{c}_{1\downarrow}^{\dagger} -ii\hat{c}_{1\downarrow}^{\dagger}(-i)\hat{c}_{1\uparrow}^{\dagger} \right) \notag \\
    &=0 
\end{align}
%For the spi
%These equations impliy that  $\hat{\rho}_{\rm MF}(x) =(\hat{S}_x)_{\rm MF}(x)=(\hat{S}_z)_{\rm MF}(x)=0$.
% and $(\hat{S}_y)_{\rm MF} \neq 0$. 
%The result is consistent with the Majorana Ising property. 
We can also evaluate other orbital-dependent operators $ \sigma_i s_j$ in Table~\ref{tab:SIR-D3h}.
For instance, $\mathcal{O}_1 = \diag (\sigma_y s_y,-\sigma_y s_y )$, $\mathcal{O}_2 = \diag (\sigma_z s_y + \sigma_x s_x, \sigma_z s_y - \sigma_x s_x )$, and $\mathcal{O}_3 =\diag (\sigma_z s_y - \sigma_x s_x, \sigma_z s_y + \sigma_x s_x )$ are evaluated as
%
%   Next, we consider the constraint from $C_3$ symmetry. 
%   Using the relations, we calculate $\hat{S}_y$, $\hat{\mathcal{O}}_1$, $\hat{\mathcal{O}}_2$, and $\hat{\mathcal{O}}_3$ for zero modes as follows.
%   \begin{align}
%    (\hat{S}_y)_{\rm MF} &= \frac{1}{2} \left(\hat{c}_{1s}^{\dagger} (s_y)_{ss'}\hat{c}_{1s'}^{\dagger}-\hat{c}_{2s}^{\dagger} ( s_y )_{ss'}\hat{c}_{2s'}^{\dagger} \right) \notag \\
%    &=\frac{1}{2} \left(-i\hat{c}_{1\uparrow}^{\dagger}\hat{c}_{1\downarrow}^{\dagger}+i\hat{c}_{1\downarrow}^{\dagger}\hat{c}_{1\uparrow}^{\dagger}+i\hat{c}_{2\uparrow}^{\dagger}\hat{c}_{2\downarrow}^{\dagger} -i\hat{c}_{2\downarrow}^{\dagger}\hat{c}_{2\uparrow}^{\dagger} \right) \notag \\
%    &=\frac{1}{2} \left(-i\hat{c}_{1\uparrow}^{\dagger}\hat{c}_{1\downarrow}^{\dagger}+i\hat{c}_{1\downarrow}^{\dagger}\hat{c}_{1\uparrow}^{\dagger}+i(-i)\hat{c}_{1\uparrow}^{\dagger}i\hat{c}_{1\downarrow}^{\dagger} -ii\hat{c}_{1\downarrow}^{\dagger}(-i)\hat{c}_{1\uparrow}^{\dagger} \right) \notag \\
%    &=0 
%    \end{align}
\begin{align}
 (\hat{\mathcal{O}}_1)_{\rm MF} &= \frac{i}{2} \left( c_{1s}^{\dagger} (s_y)_{ss'} c_{2s}^{\dagger} + c_{2s}^{\dagger} (s_y)_{ss'} c_{1s}^{\dagger}\right) \notag \\
  &= \frac{i}{2} \left( -ic_{1\uparrow}^{\dagger}c_{2\downarrow}^{\dagger}+ic_{1\downarrow}^{\dagger}  c_{2\uparrow}^{\dagger} -ic_{2\uparrow}^{\dagger} c_{1\downarrow}^{\dagger} +ic_{2\downarrow}^{\dagger} c_{1\uparrow}^{\dagger}\right) \notag \\
  &= \frac{i}{2} \left(-ic_{1\uparrow}^{\dagger}c_{2\downarrow}^{\dagger}+ic_{1\downarrow}^{\dagger}  c_{2\uparrow}^{\dagger} -i(-i)c_{1\uparrow}^{\dagger}(-i) c_{2\downarrow}^{\dagger} +iic_{1\downarrow}^{\dagger} ic_{2\uparrow}^{\dagger}\right) \notag \\
  &=0, \notag\\
(\hat{\mathcal{O}}_2)_{\rm MF} &= \frac{1}{2}\left[\left(\hat{c}_{1s}^{\dagger} (s_y)_{ss'} \hat{c}_{1s'}^{\dagger} + \hat{c}_{2s}^{\dagger} (s_y)_{ss'} \hat{c}_{2s'}^{\dagger}\right) + \left(-\hat{c}_{1s}^{\dagger} (s_x)_{ss'} \hat{c}_{2s'}^{\dagger} + \hat{c}_{2s}^{\dagger} (s_x)_{ss'} \hat{c}_{1s'}^{\dagger}\right) \right] \notag \\
   & = \left[ \left(-i\hat{c}_{1\uparrow}^{\dagger} \hat{c}_{1\downarrow}^{\dagger} + i \hat{c}_{1\downarrow}^{\dagger} \hat{c}_{1\uparrow}^{\dagger}\right) + \left(-\hat{c}_{1\uparrow}^{\dagger} \hat{c}_{2\downarrow}^{\dagger} - \hat{c}_{1\downarrow}^{\dagger} \hat{c}_{2\uparrow}^{\dagger}\right) \right] \notag \\ 
   &  = \left[ \left(-i\hat{c}_{1\uparrow}^{\dagger} \hat{c}_{1\downarrow}^{\dagger} + i \hat{c}_{1\downarrow}^{\dagger} \hat{c}_{1\uparrow}^{\dagger}\right) + \left(-i\hat{c}_{1\uparrow}^{\dagger} \hat{c}_{1\downarrow}^{\dagger} +i \hat{c}_{1\downarrow}^{\dagger} \hat{c}_{1\uparrow}^{\dagger}\right) \right] \notag \\
    & \neq 0, \notag\\
    (\hat{\mathcal{O}}_3)_{\rm MF} &= \frac{1}{2}\left[\left(\hat{c}_{1s}^{\dagger} (s_y)_{ss'} \hat{c}_{1s'}^{\dagger} + \hat{c}_{2s}^{\dagger} (s_y)_{ss'} \hat{c}_{2s'}^{\dagger}\right) - \left(-\hat{c}_{1s}^{\dagger} (s_x)_{ss'} \hat{c}_{2s'}^{\dagger} + \hat{c}_{2s}^{\dagger} (s_x)_{ss'} \hat{c}_{1s'}^{\dagger}\right) \right] \notag \\ 
   &  = \left[ \left(-i\hat{c}_{1\uparrow}^{\dagger} \hat{c}_{1\downarrow}^{\dagger} + i \hat{c}_{1\downarrow}^{\dagger} \hat{c}_{1\uparrow}^{\dagger}\right) - \left(-\hat{c}_{1\uparrow}^{\dagger} \hat{c}_{2\downarrow}^{\dagger} - \hat{c}_{1\downarrow}^{\dagger} \hat{c}_{2\uparrow}^{\dagger}\right) \right] \notag \\ 
   &  = \left[ \left(-i\hat{c}_{1\uparrow}^{\dagger} \hat{c}_{1\downarrow}^{\dagger} + i \hat{c}_{1\downarrow}^{\dagger} \hat{c}_{1\uparrow}^{\dagger}\right) - \left(-i\hat{c}_{1\uparrow}^{\dagger} \hat{c}_{1\downarrow}^{\dagger} +i \hat{c}_{1\downarrow}^{\dagger} \hat{c}_{1\uparrow}^{\dagger}\right) \right] \notag \\
    &  =0.
   \end{align}
Actually,  we find that only $(\hat{\mathcal{O}}_2)_{\rm MF}$ 
do not vanish among all possible local operators.
Note that $(\hat{\mathcal{O}}_2)_{\rm MF}$ is odd under TR and belongs to the $A_2''$ IR of $D_{3h}$, and  $B_y^3-3B_yB_x^2$ has the same symmetry properties. Thus, the lowest order coupling between the MF and magnetic feilds is $(\hat{\mathcal{O}_2})_{\rm MF}(B_y^3-3B_yB_x^2)$, which gives a magnetic octupole response.  
 
\begin{flushleft} 
{\bf S5. Topological surface states of the $A_1$ state in half-Heusler superconductors }
\end{flushleft} 

\begin{figure}[tbp]
\centering
 \includegraphics[width=16cm]{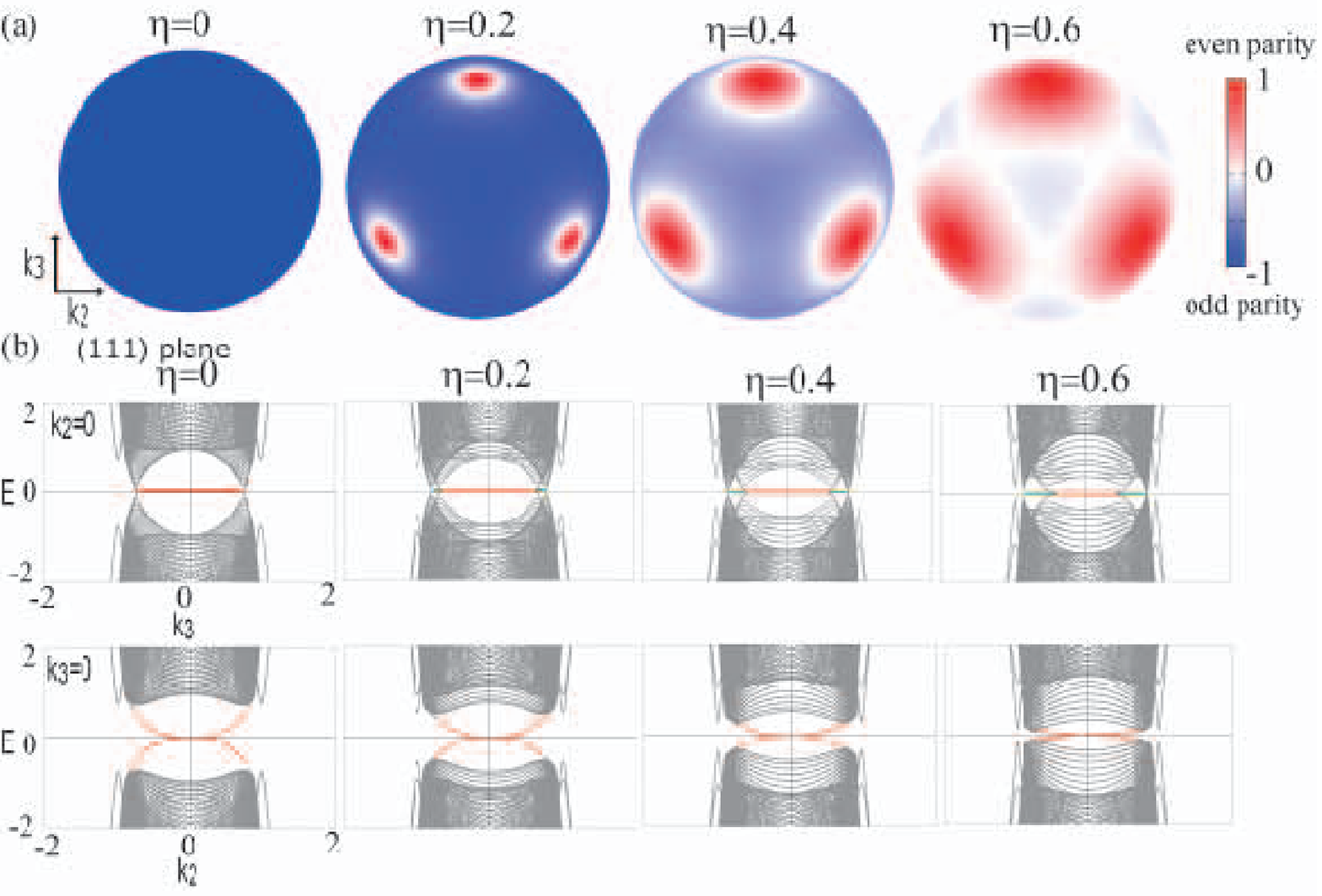}
 \caption{(a) Node structure in the Hamiltonian (\ref{eq:LK-Hami}) with parameter $(\alpha,\beta,\gamma,\delta,\mu,\Delta)=(20,-15,-15,0,-10,1)$. The color shows $(\psi^2-|\bm{d}|^2)/(\psi^2+|\bm{d}|^2)$, where $\psi$ and $\bm{d}$ describes spin-singlet and pseudospin-triplet components on the basis projected into the spin $3/2$ band~\cite{Brydon16}. The point nodes inflate into the line nodes depicted by the while lines in an increase in $\eta$. (b) Surface energy spectrum along the $k_3=\bm{k} \cdot \bm{n}_3$ (the top panels) and the $k_2=\bm{k} \cdot \bm{n}_2$ axes (the bottom panels), where  $\bm{n}_2 = \frac{1}{\sqrt{2}} (1,-1,0)$ and $\bm{n}_3 = \frac{1}{\sqrt{6}} (1,1,-2)$. The red and green lines exhibit the Majorana arc state terminating at a pair of point nodes and the Majorana flat band coming from the line nodes, respectively. For a small $\eta$, the helical MF located at $k_2=k_3=0$ still survives.}\label{fig:hHmodel}
\end{figure}

We start with the BdG Hamiltonian of the spin-3/2 half-Heusler superconductor~\cite{Brydon16}
 %\begin{align}
% \mathcal{H} = \sum_{\bm{k}} (c_{\alpha,\bm{k}}^{\dagger}, c_{\alpha,-\bm{k}}) H(\bm{k})_{\alpha \alpha'} \left( \begin{array}{@{\,} c @{\,}} c_{\alpha,\bm{k}} \\c_{\alpha,-\bm{k}}^{\dagger} \end{array} \right)
% \end{align}
 \begin{align}
  \mathcal{H}(\bm{k}) = \begin{pmatrix} H_{\rm LK} (\bm{k})-\mu & \Delta (\bm{k}) \\ \Delta (\bm{k})^{\dagger} & -H_{\rm LK} (-\bm{k})^t+\mu \end{pmatrix}, \label{eq:LK-Hami}
\end{align} 
with
\begin{align}
&H_{\rm LK}(\bm{k}) = \alpha \bm{k}^2 + \beta \sum_{i} k_i J_i^2 + \gamma \sum_{i \neq j} k_i k_j J_i J_j  + \delta \sum_{i}k_i(J_{i+1}J_i J_{i+1}-J_{i+2}J_{i}J_{i+2}), \label{eq:LKmodel} \\
&\Delta (\bm{k}) = \Delta/{\sqrt{1+\eta^2}}[ \eta 1_4 +  \sum_{i} k_i (J_{i+1}J_i J_{i+1}-J_{i+2}J_{i}J_{i+2})](e^{-i J_y \pi}). 
 \end{align}
Here $J_i$'s are the three spin matrices of $3/2$-fermions, described as 
\begin{align}
 J_x = \frac{1}{2}\begin{pmatrix} 0 & \sqrt{3} & 0 & 0 \\ \sqrt{3} & 0 & 2 & 0 \\0 & 2 & 0 & \sqrt{3} \\ 0 & 0 & \sqrt{3} & 0\end{pmatrix},  \ \ J_y = \frac{i}{2}\begin{pmatrix} 0 & -\sqrt{3} & 0 & 0 \\ \sqrt{3} & 0 & -2 & 0 \\0 & 2 & 0 & -\sqrt{3} \\ 0 & 0 & \sqrt{3} & 0\end{pmatrix}, \ \  J_z = \frac{1}{2}\begin{pmatrix} 3 & 0 & 0 & 0 \\ 0 & 1 & 0 & 0 \\0 & 0 & -1 & 0 \\ 0 & 0 & 0 & -3\end{pmatrix}.
\end{align}
The spin-orbit interactions (SOIs) proportional to $\beta$ and $\gamma$ are inversion symmetric, while the $\delta$ term is odd under inversion, describing an anti-symmetric SOI. 
The band spectrum of the normal Hamiltonian $H_{\rm LK}({\bm k})$ shows four-fold degeneracy of $J=3/2$ at the $\Gamma$ point.  Apart from the $\Gamma$ point, the energy dispersion splits into two doubly degenerate bands when $\delta =0$. When the $\delta$ term is turned on, its degeneracy splits due to the breaking of inversion symmetry. 
PH and TR operations for the BdG Hamiltonian are $\mathcal{C} = \tau_x K$ and $\mathcal{T}=e^{-i J_y \pi } K$, respectively, and  $T_d$ for the BdG Hamiltonian is generated by rotations 
 $\mathcal{U}_{q,\bm{n}} =\diag [e^{i\frac{i 2\pi}{q} \bm{J} \cdot \bm{n}},e^{-i\frac{i 2\pi}{q} \bm{J}^* \cdot \bm{n}}]$ and mirror reflections $\tilde{\sigma}=\diag[\sigma, \sigma^*]$ where $\sigma$ represents diagonal and vertical mirror reflections for the normal Hamiltonian. 

In Fig.~\ref{fig:hHmodel} (a), we show the node structure of the superconducting state on the spherical Fermi surface. 
When  $\eta = 0$, $\Delta(\bm{k})$ is the pure septet pairing and hosts point nodes on the $x$, $y$, and $z$ axes. For a small $\eta$, the mixture of spin-singlet and spin-septet components inflates point nodes to line nodes in a similar way to other non-centrosymmetric SCs~\cite{Bauer12}. Crucially, in the $[111]$ direction, the spin-septet component is most dominant on the Fermi surface.
In Fig.~\ref{fig:hHmodel} (b), we show topological surface states of the system, which are calculated by replacing $k_i, k_i^2$ with $\sin k_i, 2(1-\cos k_i)$ and imposing the open boundary condition along the $[111]$ direction. The obtained surface energy spectra are shown. When $\eta=0$, the pure spin-septet superconductor hosts six point nodes, which induce the Majorana arc state in the $k_3$ direction, which connects projected point nodes on the surface BZ. 
Adding the spin-singlet component ($\eta \neq 0$), line nodes arise and induce non-degenerate surface flat bands. See the top panel of Fig.~\ref{fig:hHmodel} (b). Despite the mixing of the spin-singlet component, the helical MF located at $\bm{k}_{\parallel}=0$ is stable when $\eta < \eta_c \simeq 0.9$. The similar topological surface states are obtained when $\beta \neq \gamma$ and $\delta \neq 0$.  For Fig. 2 in the main paragraph, we take the parameters as $(\alpha,\beta, \gamma, \delta, \mu, \Delta, \eta)=(20,-15,-10,1,-10,0.5,0.2)$, where we have confirmed the presence of the Majorana arc states and the non-degenerate surface flat bands on the $(111)$ surface.

The existence of the Majorana arc states is ensured by  
the magnetic 1D winding number
 \begin{align}
W(\bm{k}_{\parallel},\Gamma_{\sigma}) \equiv \frac{i}{4\pi} \int^{\pi}_{-\pi} d k_{\perp} \; \trace \left[ \Gamma_{\sigma} \mathcal{H}(\bm{k})^{-1} \partial_{k_{\perp}} \mathcal{H}(\bm{k})\right], 
\label{eq:1dwinding}
\end{align}
where $\Gamma_{\sigma} = \tilde{\sigma} \mathcal{T} \mathcal{C}$ with $\tilde{\sigma}$ the diagoal mirror reflection.
Here $(\bm{k}_{\parallel}, k_{\perp})$ are the momentum parallel to and normal to the surface, respectively, and ${\bm k}_{\parallel}$ in the left hand side of Eq.(\ref{eq:1dwinding}) should be on the diagonal mirror plane, say ${\bm k}_{\parallel}=(0,k_3)$. 
%
%Here we choose the coordinate system $(k_{\perp}, {\bm k}_{\parallel})=(k_1, k_2, k_3)$ in the momentum space, where $k_a=\bm{k} \cdot \bm{n}_{a}$ with $\bm{n}_1 = \frac{1}{\sqrt{3}} (1,1,1)$, $\bm{n}_2 = \frac{1}{\sqrt{2}} (1,-1,0)$, and $\bm{n}_3 = \frac{1}{\sqrt{6}} (1,1,-2)$. 
%
%
%First, we clarify the symmetry-protected topological numbers in the BdG Hamiltonian. When $\eta=0$, the system hosts point nodes on the $x$, $y$, and $z$ axes.
%
% induce Majorana arc states in the mirror planes. On the ($111$) plane, $\Gamma_{\sigma}$ and $\bm{k}_{\parallel}$ are concretely described as $\Gamma_{\sigma}=\mathcal{U}_{2,\bm{n}_2} \mathcal{T} \mathcal{C}$ and $\bm{k}_{\parallel} =(k_2,k_3)= (\bm{k} \cdot \bm{n}_{2}, \bm{k} \cdot \bm{n}_{3})$, where   $\bm{n}_1 = \frac{1}{\sqrt{3}} (1,1,1)$, $\bm{n}_2 = \frac{1}{\sqrt{2}} (1,-1,0)$, and $\bm{n}_3 = \frac{1}{\sqrt{6}} (1,1,-2)$. 
Because of $C_3$ symmetry of the surface, we have Majorana arcs on three equivalent directions.
%from which we have a distorted helical MF with $C_{3v}$ symmetry. 
%the following three  magnetic chiral operators, 
%$\Gamma_{\sigma}$,  $\mathcal{U}_{3,\bm{n}_1}^{\dagger} \Gamma_{\sigma} \mathcal{U}_{3,\bm{n}_1}$ and $(\mathcal{U}_{3,\bm{n}_1}^{\dagger})^2 \Gamma_{\sigma} (\mathcal{U}_{3,\bm{n}_1})^2$ give the same winding number. 
In particular, the three Majorana arcs form a single distorted helical MF with $C_{3v}$ symmetry centered $\bm{k}_{\parallel}=0$. 
When $\eta \neq 0$, the mixture between the spin-singlet and spin-septet components gives rise to line nodes, as mentioned in the above.
The line nodes host the 1D winding number $W (\bm{k}_{\parallel}, \Gamma)$ of the conventional chiral operator $\Gamma = -i \mathcal{T} \mathcal{C}$~\cite{Sato11}, and thus 
the non-degenerate flat bands mentioned above appear.
% protected by 
%with the conventional chiral operator 
%appear along with the line nodes. 
We emphasize here that the Majorana arc states by $W(\bm{k}_{\parallel},\Gamma_{\sigma}) $ and non-degenerate flat bands by $W(\bm{k}_{\parallel},\Gamma) $ coexist and the helical MF survives as long as $\eta < \eta_c$, where $\eta_c$ is a topological phase transition point with respect to  $W (\bm{k}_{\parallel}=0, \Gamma_{\sigma})$.  

%In the following, we demonstrate the nodal structure and the topological surface states in a situation that the spin-$3/2$ ($1/2$) band curves downward (upward) and the spherical symmetric case, i.e., $\beta = \gamma$ and $\delta=0$, which simplification does not alter topological properties.  

\begin{flushleft} 
{\bf S6. Energy gap and magnetic octupole response}
\end{flushleft} 

To see the magnitude of the energy gap for the helical MF under the magnetic octupole response, we consider Eq.~(\ref{eq:LKmodel}) with the Zeeman magnetic field: $H_{LK}(\bm{k}) +\mu \bm{B} \cdot \bm{J}$. On the basis that diagonalizes $H_{\rm LK}(\bm{k})$ with $\delta = \mu =0$, the Hamiltonian on the $[111]$ direction becomes 
\begin{align}
 U^{\dagger} H_{\rm LK} (k) U = \frac{k^2}{4}\diag [4\alpha + 5\beta + 4\gamma,4\alpha + 5\beta + 4\gamma,4\alpha + 5\beta - 4\gamma,4\alpha + 5\beta - 4\gamma], \label{eq:diagLKmodel}
\end{align} 
with
\begin{align}
 U= \frac{e^{i\frac{\pi}{4}}}{\sqrt{6}} \begin{pmatrix} -1-i & -i & 0 & \sqrt{3} \\ -i \sqrt{3}  & 0 &1 &-1-i \\ 0 & \sqrt{3} & -1-i&-i \\ 1 & -1-i &-i \sqrt{3} &0\end{pmatrix}.
\end{align}
The TR operator and the $C_{3v}$ symmetry operator are also given by
\begin{align}
&U^{\dagger} e^{-i J_y \pi} U^{\ast} = \begin{pmatrix} 0& 1 & 0 &0 \\-1 & 0 & 0&0 \\ 0 & 0& 0&1 \\ 0 & 0 & -1&0\end{pmatrix}, \label{eq:diagT} \\
&U^{\dagger} e^{-i \frac{J_x+J_y+J_z}{\sqrt{3}} \frac{2\pi}{3} } U = \begin{pmatrix} -1& 0 & 0 &0 \\0 & -1 & 0&0 \\ 0 & 0& \frac{1+i}{2}&\frac{1-i}{2} \\ 0 &0& -\frac{1+i}{2} & \frac{1-i}{2} \end{pmatrix}, \label{eq:diagC3} \\
&U^{\dagger} e^{-i \frac{J_x-J_y}{\sqrt{2}} \pi } U = \begin{pmatrix} 0& -e^{i\frac{\pi}{4}} & 0 &0 \\-e^{i\frac{3\pi}{4}} & 0 & 0&0 \\ 0 & 0& 0&e^{i\frac{\pi}{4}} \\ 0 &0& e^{i\frac{3\pi}{4}} & 0 \end{pmatrix}. \label{eq:diagC2}
\end{align}
Projecting the Hamiltonian into the  spin-$3/2$ band and including the anti-symmetry SOI and the magnetic Zeeman term as a perturbation up to the third order, Eqs.~(\ref{eq:diagLKmodel}), (\ref{eq:diagT}), (\ref{eq:diagC3}), and (\ref{eq:diagC2}) are given by 
\begin{align}
 H_{\rm eff}(k) &= \frac{k^2}{4} (4\alpha + 5\beta + 4\gamma) \sigma_0 - \frac{\sqrt{3}k}{2} \delta (\sigma_x + \sigma_y) + \frac{\sqrt{3}}{2} \mu B_1 (\sigma_x +\sigma_y +\sigma_z)   \notag \\
  & + \frac{3}{2\Delta \epsilon(k)} \mu^2 (B_2^2 +B_3^2) \sigma_0 + \frac{\sqrt{3}}{8\Delta \epsilon(k)^2} \mu^3 \Big[ \left\{-2B_1 (B_2^2 + B_3^2) + 2\sqrt{2}B_3 (B_3^2-3B_2^2) \right\}\sigma_z \notag \\
  &-\left\{ 2 B_1 (B_2^2+B_3^2) +\sqrt{2} B_3 (B_3^2 - 3 B_2^2) +\sqrt{3} B_2 (B_2^2-3B_3^2) \right\} \sigma_x  \notag \\
  &- \left\{ 2 B_1 (B_2^2+B_3^2) +\sqrt{2} B_3 (B_3^2 - 3 B_2^2)- \sqrt{3} B_2 (B_2^2-3B_3^2) \right\} \sigma_y \Big], \label{eq:effmodel}\\
  T_{\rm eff} &= i \sigma_y K, \label{eq:effT} \\
  U_{3,{\rm eff}} &= - \sigma_0, \label{eq:effC3}\\
  U_{\sigma_v,{\rm eff}} &= \frac{i}{\sqrt{2}}(\sigma_x - \sigma_y). \label{eq:effC2}
\end{align} 
where ($\sigma_0, \bm{\sigma}$) are the $2 \times 2$ identity matrix and the Pauli matrices, $\Delta \epsilon(k) \equiv \epsilon_{3/2}(k)-\epsilon_{1/2}(k)$, and $(B_1,B_2,B_3) \equiv (\frac{1}{\sqrt{3}} \{B_x + B_y +B_z\}, \frac{1}{\sqrt{2}}\{B_x-B_y\}, \frac{1}{\sqrt{6}}\{B_x+B_y-2B_z\})$. Introducing $\sigma_1 \equiv \frac{1}{\sqrt{3}} (\sigma_x + \sigma_y+\sigma_z)$, $\sigma_2 \equiv \frac{1}{\sqrt{2}} (\sigma_x - \sigma_y)$, and $\sigma_3 \equiv \frac{1}{\sqrt{6}} (\sigma_x + \sigma_y-2 \sigma_z)$, Eq.~(\ref{eq:effmodel}) is recast into
\begin{align}
 H_{\rm eff}(k) &= \frac{k^2}{4} (4\alpha + 5\beta + 4\gamma) \sigma_0 - k \delta \left( \sigma_1 + \frac{1}{\sqrt{2}} \sigma_3 \right) + \frac{3}{2} \mu  B_1 \sigma_1  + \frac{3}{2\Delta \epsilon(k)} \mu^2 (B_2^2 +B_3^2) \sigma_0   \notag \\
  &- \frac{3}{4\Delta \epsilon(k)^2} \mu^3 \Big[ B_1(B_2^2+B_3^2) \sigma_1 + \frac{1}{\sqrt{2}} B_2 (B_2^2 -3 B_3^2) \sigma_2+B_3(B_3^2-3 B_2^2) \sigma_3 \Big]. \label{eq:effmodelv2}
\end{align}

\begin{figure}[tbp]
\centering
 \includegraphics[width=12cm]{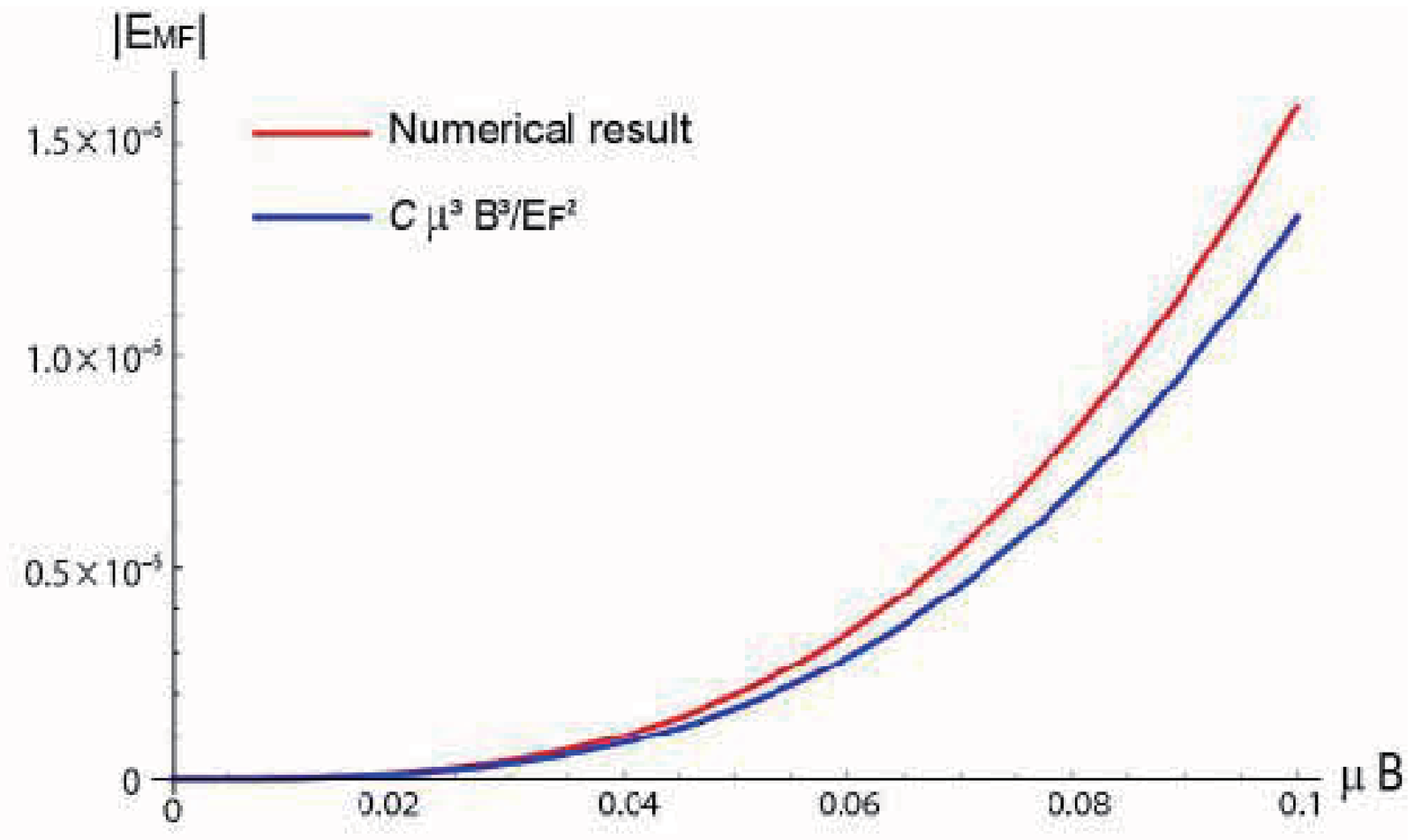}
 \caption{The magnitude of the energy gap $E_{\rm MF}$ as a function of $B$. The red thick and blue thick lines represent the numerical and analytical results, respectively, where $C=3 \sqrt{2}/32$. The parameters and the direction of the Zeeman magnetic field are chosen as $(\alpha,\beta,\gamma,\delta,\mu,\Delta,\eta)=(20,-15,-15,0,-10,1,0)$ and $(\theta,\phi)=(\pi/2,0)$, respectively.  }\label{fig:EMKPgap}
\end{figure}

Within the effective Hamiltonian, the BdG Hamiltonian is given by
\begin{align}
 \mathcal{H}_{\rm eff}(k) = \begin{pmatrix} H_{\rm eff } (k) -E_{\rm F} \sigma_0 & \Delta(k) \\ \Delta(k)^{\dagger} & -H_{\rm eff}(-k)^t + E_{\rm F} \sigma_0 \end{pmatrix},
\end{align}
with the Fermi energy $E_{\rm F}$ and the gap function $\Delta (k) = \frac{\Delta}{\sqrt{1+\eta^2}} \left\{\eta \sigma_0 - \frac{\sqrt{3}k}{2} (\sigma_x+\sigma_y) \right\}i \sigma_y $. In addition,  Eqs.~(\ref{eq:effT}), (\ref{eq:effC3}), and (\ref{eq:effC2}) are extended into the Nambu space:
\begin{align}
 &\mathcal{T}_{\rm eff} = i \sigma_y \tau_0 K, \\
 &\mathcal{U}_{3, {\rm eff}} = - \sigma_0 \tau_0, \\
  &\mathcal{U}_{\sigma, {\rm eff}} = \frac{i}{\sqrt{2}} (\sigma_x \tau_z - \sigma_y \tau_0) ,
\end{align}
where $(\tau_0,\bm{\tau})$ are the $2 \times 2$ identity matrix and the Pauli matrices describing the Nambu space. The BdG Hamiltonian hosts the magnetic chiral symmetry $\{\mathcal{H}_{\rm eff}(k) , \Gamma_{\sigma}\}=0$ with $\Gamma_{\sigma} \equiv \mathcal{U}_{\sigma_v, {\rm eff}} \mathcal{T}_{\rm eff} \mathcal{C} $, which involves a nonzero 1D winding number and ensures the MF on the surface. Here $\mathcal{C}=\tau_x K$ is the PH operator. Furthermore, due to the $C_3$ symmetry, we have different magnetic chiral operators: $\mathcal{U}_{3,{\rm eff}}^{\dagger} \Gamma_{\sigma} \mathcal{U}_{3,{\rm eff}}$ and $(\mathcal{U}_{3,{\rm eff}}^{\dagger})^2 \Gamma_{\sigma} (\mathcal{U}_{3,{\rm eff}})^2$, which also protect the MF at the same time.  Hence, the corresponding MF can be a quantum state $| u_0^{(a)} \rangle$ ($a=1,2$) that is an simultaneous eigenstate of $\Gamma_{\sigma}= \frac{1}{\sqrt{2}} (\sigma_z \tau_y + \sigma_0 \tau_x)$ and $\mathcal{U}_{3, {\rm  eff}} $, e.g., $\Gamma_{\sigma } | u_0^{(a)} \rangle = | u_0^{(a)} \rangle$ and $\mathcal{U}_{3,{\rm eff}}| u_0^{(a)} \rangle = -| u_0^{(a)} \rangle$. Taking the symmetry constraints into account, we calculate the expectation value $\langle u_0^{(a)} | \tilde{\sigma}_i| u_0^{(b)} \rangle$ with $\tilde{\sigma}_i =\diag [\sigma_i, -\sigma_i^{\ast}]$ ($i=0,1,2,3$). After algebraic calculations, we find
\begin{align}
  \{\Gamma_{\sigma_v},\tilde{\sigma}_0\}=\{\Gamma_{\sigma_v},\tilde{\sigma}_1\}=[\Gamma_{\sigma_v},\tilde{\sigma}_2]=\{\Gamma_{\sigma_v},\tilde{\sigma}_3\}=0,
\end{align}
so only $\langle u_0^{(a)} | \tilde{\sigma}_2| u_0^{(b)} \rangle$ remains nonzero. The Zeeman magnetic term that affects the MF becomes
\begin{align}
\langle u_0^{(a)} | \tilde{H}_{\rm eff, Zeeman }| u_0^{(b)} \rangle = - \frac{3\sqrt{2}}{8\Delta \epsilon(k)^2}  \mu^3 B_2 (B_2^2 -3 B_3^2) \langle u_0^{(a)} | \tilde{\sigma}_2| u_0^{(b)} \rangle.
\end{align}
When the spin-$3/2$ ($1/2$) band curves oppositely and $\delta \ll 1$, $\Delta \epsilon(k) \sim 2 E_{\rm F}$. Then, the magnitude of the energy gap is estimated as 
\begin{align}
|E_{\rm MF}| \propto \frac{3 \sqrt{2}}{32} \frac{\mu^3 B^3}{E_{\rm F}^2} \sin^3 \theta \cos 3\phi ,
\end{align}
 when $(B_1,B_2,B_3)=B(\cos\theta,\sin \theta \cos \phi, \sin \theta \sin \phi)$. The comparison between the analytical and the numerical results are shown in Fig.~\ref{fig:EMKPgap}.
%\bibliography{multipole}

\end{document}